\documentclass[aps,prb,twocolumn,reprint]{revtex4-2}

\usepackage{amsmath,amsfonts}
\usepackage{mathtools}

\usepackage{siunitx}

\usepackage[pdftex]{graphicx}
\usepackage[whole,autotilde]{bxcjkjatype} 

\usepackage{here}

\usepackage[caption=false]{subfig} 

\usepackage[colorlinks,allcolors=blue]{hyperref}

\usepackage[final,marginclue,inline,author=]{fixme}
\fxsetup{theme=color,marginface=\small,inlineface=\small}

\graphicspath{{./Figs2311/}}

\begin{document}

\title{Quantum coherent control of linear and nonlinear thermoelectricity on graphene nanostructure heat engines}
 
\author{Yuga Kodama and Nobuhiko Taniguchi}
\email{taniguchi.n.gf@u.tsukuba.ac.jp}
\affiliation{Physics Division, Faculty of Pure and Applied Sciences,
  University of Tsukuba, Tennodai Tsukuba 305-8571, Japan} 

\date{\today}

\begin{abstract}
We theoretically show how structural modifications and controlling quantum coherency can enhance linear and nonlinear thermoelectric performance in graphene nanostructure heat engines. Although graphene has emerged as a promising material for a nanoscale heat engine due to its high coherency and tunable electronic properties, its large lattice thermal transport often limits its thermal efficiency. Using the density-functional tight-binding method, we demonstrate that one can suppress lattice thermal transport, degrading the thermal efficiency by deliberately manipulating the junction's bending angle at low temperatures. We further argue that applying an optimal local gate voltage unleashes its great potential in achieving excellent efficiency and reasonably high output power that persist in the fully nonlinear regime. 

\end{abstract}


\maketitle


\section{introduction}

%


Over the last decade, low-dimensional and nanoscale materials have attracted much attention as promising candidates for a thermoelectric engine that can directly convert heat into electric power~\cite{Dresselhaus07}. Sharp resonances due to discrete levels in a nanoscale system naturally arise an energy filtering effect, which makes the system act as a heat engine by exchanging particles between external reservoirs. 
One typically assesses thermoelectric performance by the linear-response quantity called the figure of merit, $ZT = G S^{2} T / \kappa$, which reflects temperature $T$, conductance $G$, Seebeck coefficient $S$, and thermal conductance $\kappa$. A higher value indicates greater thermal efficiency. Researchers have long recognized that materials with the density of states (DOS) characterized by sharp peaks and acute changes can yield a high value of $ZT$, making nanoscale materials a viable option for improved thermoelectric performance~\cite{Hicks93,Hicks93b,Mahan96}. 

Nanoscale systems have a further advantage of greater control in designing and engineering the structure. A nanostructure maintains quantum coherence over the system, and its transport accordingly depends strongly on junction types in contrast to bulk materials. 
Thermoelectric phenomena in nanoscale systems often appear as nonlinear quantum transport~\cite{Sanchez16,Benenti17}. Several theoretical bounds in the nonlinear thermoelectric processes have been discussed~\cite{Whitney13,Whitney14,Yamamoto15}. 
For a given nanoscale system, it is worthwhile to ask what kinds of minor structural modifications can enhance thermoelectric performance effectively. Such insights will be highly beneficial to advance thermoelectric technology and nanoscale heat engines. One successful approach is to exploit quantum coherence and destructive interference. Studies have shown that enhancement of the thermal efficiency occurs when the transmission is significantly lowered~\cite{Karlstrom11,Lambert16,Bergfield09,Bergfield10,Abbout13} or occurs near the Fano resonance~\cite{Finch09,Gomez-Silva12,Trocha12,Garcia-Suarez13,Bevilacqua16,Wojcik16,Taniguchi20}. The latter is particularly appealing because the effect seems to persist in the fully nonlinear regime~\cite{Taniguchi20}, where one usually operates nanoscale heat engines.

Graphene nanoribbons have great potential as nanoscale thermoelectric material due to their high phase coherency and tunable electronic properties~(see \cite{SaitoBook98,KatsnelsonBook12} for general properties). One can utilize its versatile structures to control ballistic transport. Graphene, however, has large lattice thermal transport that often worsens thermal efficiency. Therefore it is crucial to suppress phonon transport. Extensive research has been conducted on thermoelectric properties of graphene nanoribbons, with considerable efforts to identify favorable structures that can achieve a higher value of $ZT$ by exploring changes in the width and edge orientation and whether armchair or zigzag sections~\cite{Ouyang09,Divari10,Savin10,Sevincli10,Xu10,Mazzamuto11,Zhang12b,Rosales13,Sevincli13,Li14} or exploiting electron's quantum interference in a ring geometry~\cite{Saiz-Bretin15,Saiz-Bretin19}. 

In this paper, we choose a rhombus-shaped graphene dot and theoretically demonstrate how linear and nonlinear thermoelectric performance gets significantly improved by introducing two types of structural modifications:  (1) applying the local gate voltage in the middle to make electron's transport ring-like (Fig.~1), and (2) changing bending angles at the junction (Fig.~2).  
As gate voltage impacts little on phonon transport, these two types of modification help control electron and phonon transport separately. We systematically explore which bending suppresses lattice transport most and how local gate voltage helps improve thermoelectric performance. Although phonons deteriorate thermal efficiency, we will find such a controllable quantum nanostructure produces excellent efficiency and reasonably high output power at optimized parameters, particularly at low temperatures.
The result contrasts with straight nanoribbons that are weakly thermoelectric with typically $ZT\lesssim 0.1$. 
Besides examining linear-response quantities like $ZT$, we investigate thermal efficiency and output power in the fully nonlinear regime, where a nanoscale heat engine usually operates. We also show normalized quantities enable us to estimate various nonlinear thermoelectric performance quite well from linear-response quantities.


This paper is structured as follows.  In Sec.~\ref{sec:model-and-method}, we introduce the graphene nanoribbon systems and explain structural modifications we will analyze. We will also present theoretical descriptions and numerical methodology. In Sec.~\ref{sec:linear-response}, we discuss how various types of bending at the junction affect phonon thermal conductance. After identifying a type of junction bending that reduces lattice transport, we optimize the figure of merit and linear-response thermoelectric performance by introducing the local gate voltage. Sec.~\ref{sec:nonlinear-response} devotes itself to the nonlinear thermoelectric performance of the optimized heat engines. Finally, we conclude in Sec.~\ref{sec:conclusion}. 

\section{model and method}
\label{sec:model-and-method}

%
%
%

\subsection{Structures of the model}
\label{sec:model}

\begin{figure}
  \centering
  \subfloat[\label{fig:rhombus-ring}]{
    \includegraphics[width=0.5\linewidth]{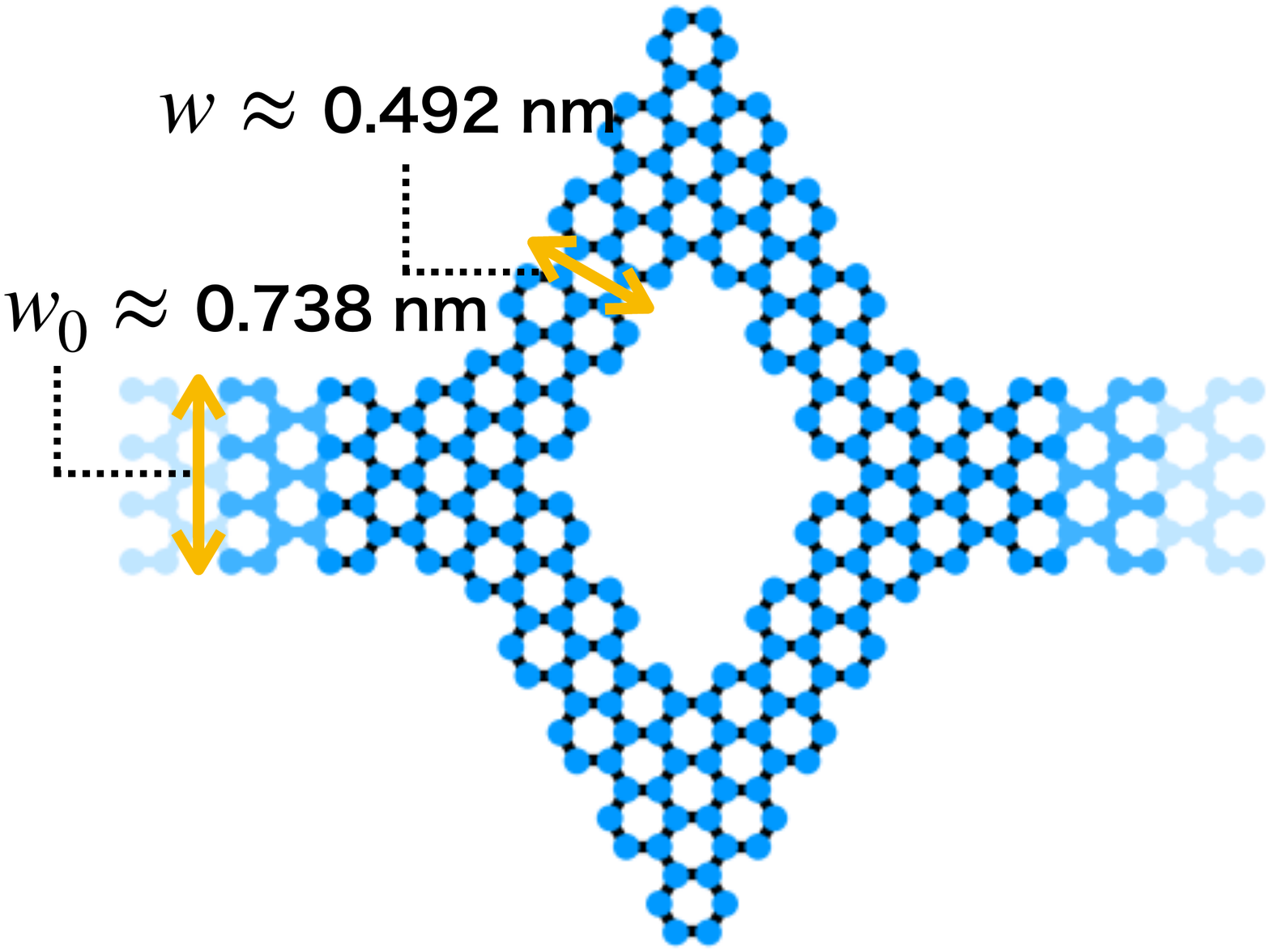}
      }
  \subfloat[\label{fig:rhombus-dot}]{
    \includegraphics[width=0.5\linewidth]{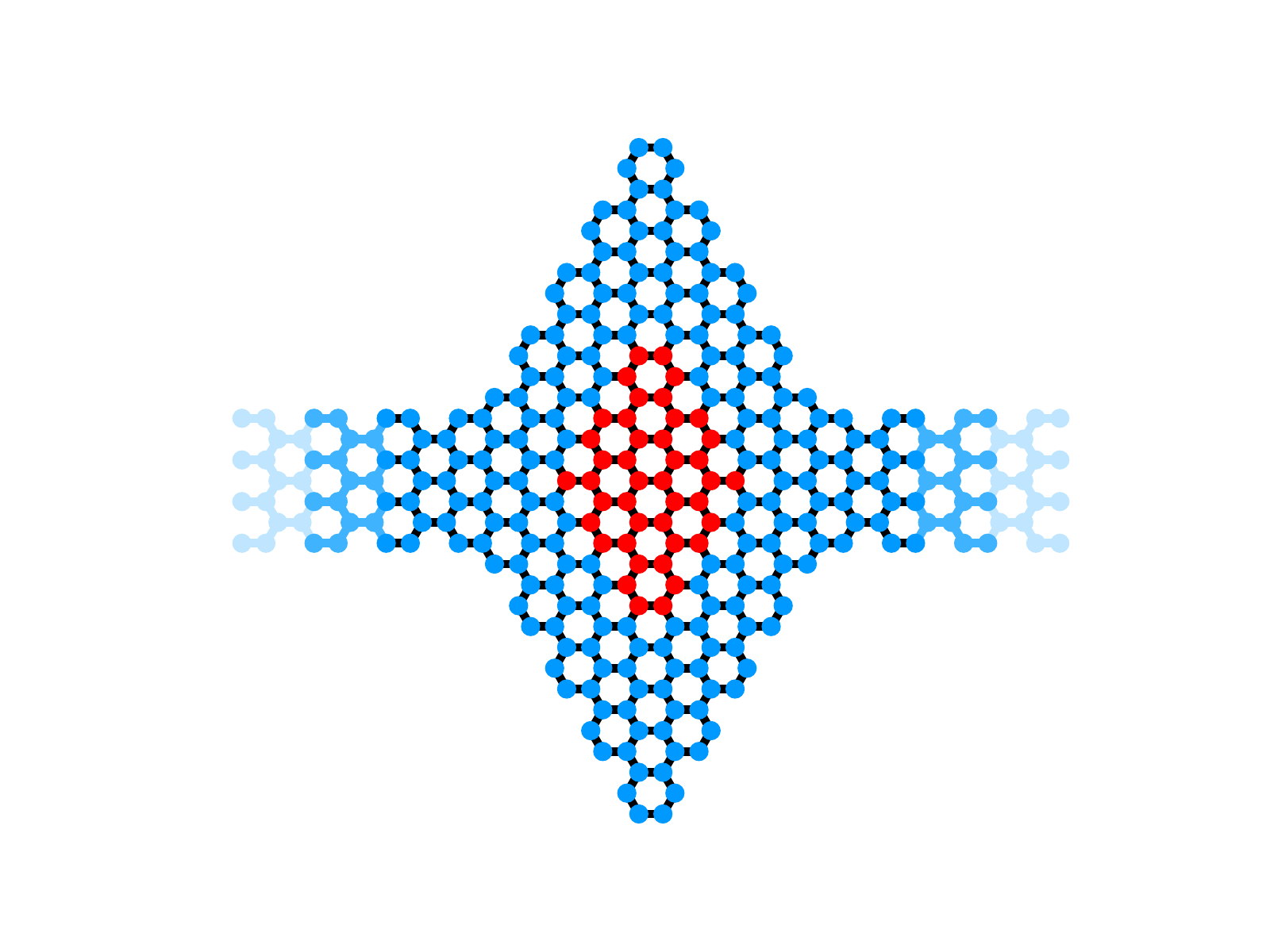}
      }
  \caption{Two types of graphene nanostructures: (a) rhombus ring, (b) rhombus dot with applying gate voltage on the red region $D$. 
  \label{fig:graphene-rhombus}}
\end{figure}

\begin{figure}[hbp]
  \centering
  \includegraphics[width=0.95\linewidth]{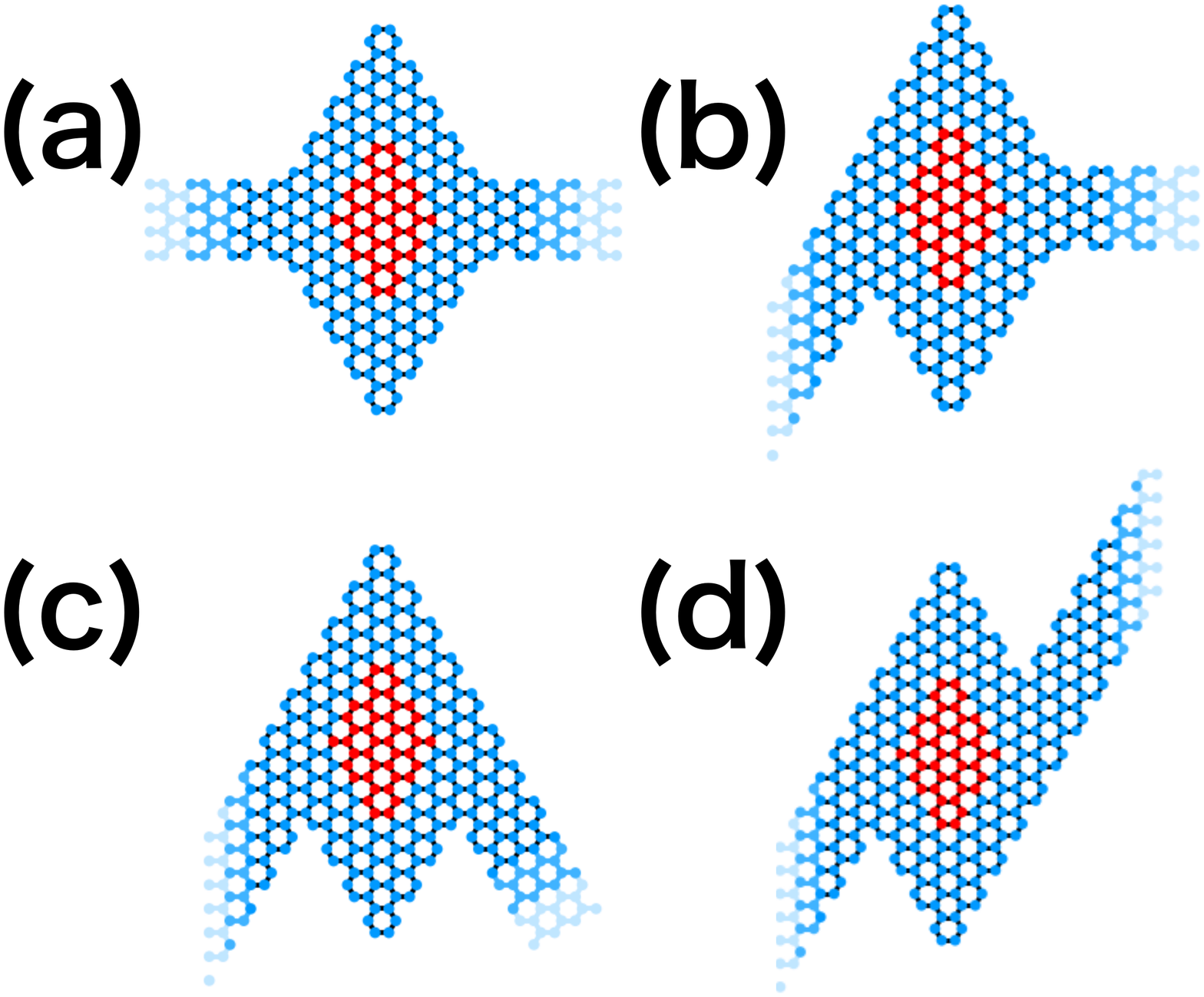}
  \caption{Four configurations of bending angles of external leads:  
  (a) simple attached, (b) soft bent, (c) hard bent, (d) double bent. The same naming scheme is applied to rhombus rings.
}
  \label{fig:four-bending-angles}
\end{figure}

\begin{figure}
  \centering
\subfloat[]{
  \includegraphics[scale=0.4]{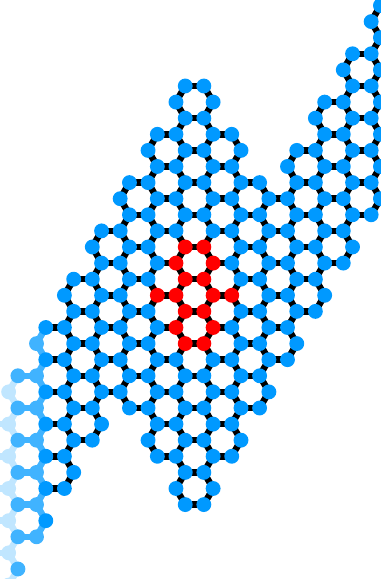}
   } \quad
\subfloat[]{
  \includegraphics[scale=0.4]{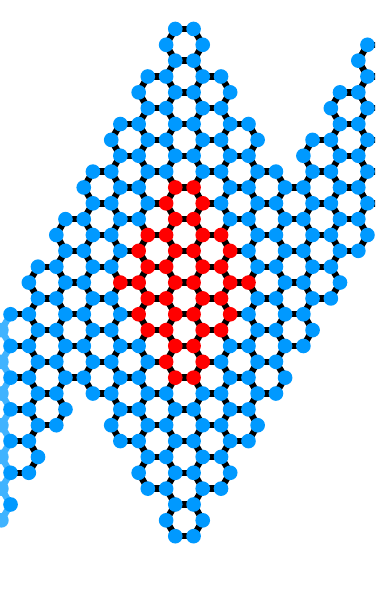}
   } \quad
\subfloat[]{
  \includegraphics[scale=0.4]{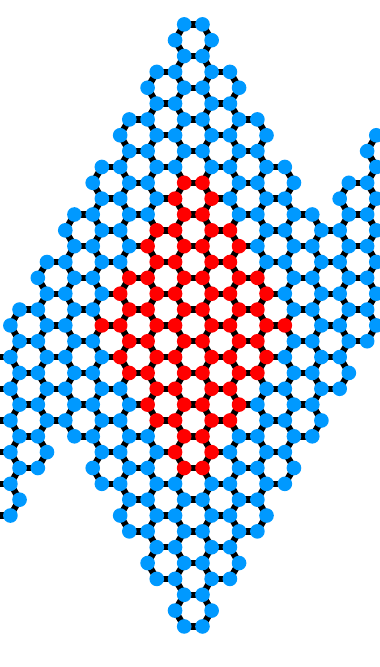}
   } \quad
\subfloat[]{
  \includegraphics[scale=0.4]{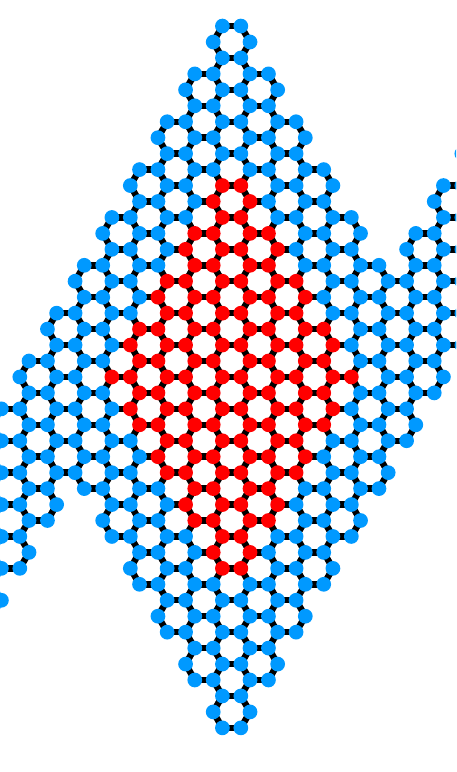}
   }
  \caption{The size $N$ of a graphene rhombus refers to the number of hexagons along its diagonal direction: (a) $N=9$, (b) $N=11$, (c) $N=13$ and (d) $N=15$. The red region is punctured for rhombus rings.}
  \label{fig:rhombus-size}
\end{figure}

We consider a graphene rhombus ring or dot (Fig.~\ref{fig:graphene-rhombus}), connecting with the two external reservoirs via two nanoribbon contacts of width $w_{0}\approx\SI{0.738}{nm}$. The width of a rhombus ring is $w\approx \SI{0.492}{nm}$, and for a rhombus dot, we apply local gate voltage in the middle region [shown as the red region in Fig.~\ref{fig:rhombus-dot}]. The presence of local gate voltage makes electron transport resemble a ring geometry, while it has little impact on phonons; we ignore its effect on phonon transport. We investigate four configurations of bending angles of the junction: (a) simple attached, (b) soft bent, (c) hard bent, and (d) double bent (Fig.~\ref{fig:four-bending-angles}).  All sections of graphene are assumed to have armchair edges except at the junction where five-membered arcs are present. In addition, to see how the size of a rhombus affects lattice transport, we choose four different types of rhombus size for each configuration (Fig.~\ref{fig:rhombus-size}). 

\subsection{Microscopic description}

We analyze linear and nonlinear thermoelectricity based on the microscopic description. Since the electron-phonon mean free path in graphene nanostructures exceeds tens of $\unit{\micro m}$ at the room temperature~\cite{Gunlycke07}, we ignore the electron-phonon interaction. The total Hamiltonian becomes $H=H_{\text{el}}+H_{\text{ph}}$, where the electron and phonon parts are given by 
\begin{align}
  H_{\rm el} &= \epsilon_{g} \sum_{i \in D}  c_{i}^{\dag} c_{i} - t \sum_{\langle i,j\rangle} \left( c_{i}^{\dag}c_{j} + c_{j}^{\dag}c_{i} \right),\\
  H_{\rm ph} &= \frac{1}{2} \sum_{i} \dot{\boldsymbol{u}}_{i}^{T} \dot{\boldsymbol{u}} +  \frac{1}{2} \sum_{i,j} \boldsymbol{u}_{i}^{T} K_{ij} \boldsymbol{u}_{j},
\end{align}
where $c_{i}^{\dagger}$ and $\boldsymbol{u}_{i}$ refer to an electron creation operator and a lattice displacement vector at the site $i$. Here we have employed the nearest-neighboring approximation for $H_{\text{el}}$, setting $t=\SI{2.8}{eV}$. Inside the region $D$ (Fig.~\ref{fig:rhombus-dot}), we introduce local gate voltage $\epsilon_{g}$, which will control the electron quantum coherence. 
%
For each structure, we have numerically obtained the force constant matrix $K_{ij}$ in $H_{\text{ph}}$ by using the density functional tight binding approach with the help of DFTB+~\cite{Hourahine20}. To do this, with Slater-Koster parameters for C and H atoms~\cite{Niehaus01}, we have employed the conjugate gradient method to achieve geometrical optimization including the reservoirs, until the inter-atomic forces become less than $\num{e-5}$ a.u.~\cite{Saiz-Bretin19}. After that, $K_{ij}$ is numerically available as the Hessian matrix of the lattice potential.
To investigate linear and nonlinear transport quantities, we evaluate the transmission of an electron, $\mathcal{T}_{\text{el}}(E)$~\cite{Meir92,HaugBook08}, and of phonon $\mathcal{T}_{\text{ph}}(E)$~\cite{Sandonas19}, using the standard technique of nonequilibrium Green's functions (see also Sec.~\ref{sec:nonlinear-thermoelectricity}). During the process, we have also used Kwant~\cite{Groth14}. 

\subsection{Nonlinear thermoelectricity}
\label{sec:nonlinear-thermoelectricity}

As a concrete realization of a nanoscale heat engine, we consider the graphene rhombus connecting via the nanoribbon contacts with the reservoirs (the left and right leads, $a=L/R$) with different electrochemical potentials $\mu_{L}<\mu_{R}$ and temperatures $T_{L}>T_{R}$. In this setting, the temperature voltage drives heat flow against the potential bias and converts heat into electric work. 
Output power $P$ and the thermal efficiency $\eta$ of this heat engine are defined by
\begin{align}
& P= (\mu_{R} - \mu_{L}) I_{L}; \quad 
\eta = \frac{P}{J_{L}},
\label{def:power-efficiency}
\end{align}
where $I_{a}$ and $J_{a}$ are particle and heat inflows from the lead $a$~(see \cite{Benenti17} for a review). Since electrons and phonons contribute to heat flow, one can express $J_{a}$ as $J_{a}^{\text{el}}+J_{a}^{\text{ph}}$. We also introduce the electron thermal efficiency $\eta_{\text{el}}=P/J_{L}^{\text{el}}$ by ignoring the phonon deterioration effect. The knowledge of transmission spectra enables us to evaluate these flows in the fully nonlinear regime \cite{Sanchez16,Benenti17} as
\begin{align}
  I_{L} &= \int_{-\infty}^{\infty} \frac{dE}{h} \mathcal{T}_{\rm el}(E) [f_{L}(E) - f_{R}(E)], \label{eq: particle flow}\\
  J^{\rm el}_{L} &=  \int_{-\infty}^{\infty} \frac{dE}{h} (E - \mu_{L})\, \mathcal{T}_{\rm el}(E) [f_{L}(E) - f_{R}(E)], \label{eq: electron heat flow}\\
  J^{\rm ph}_{L} &= \int \frac{dE}{h} E\, \mathcal{T}_{\rm ph}(\varepsilon)[n_{L}(E) - n_{R}(E)]. 
\label{eq:J-ph}
\end{align}
with Fermi distribution $f_{a}(E) = 1/[e^{\beta_{a}(E-\mu_{a})} + 1]$ and Bose distribution $n_{a}(E) = 1/[e^{\beta_{a} E} - 1]$, with the inverse temperature $\beta_{a}=1/k_{B}T_{a}$ of the lead $a$. 

Efficiency $\eta$ is bound from above by the Carnot efficiency $\eta_{C}=(T_{L}-T_{R})/T_{L}$, while the natural scale for the output power is $P_{\Delta T}=k_{B}^{2}(T_{L}-T_{R})^{2}/4h$ [see Eq.~\eqref{eq:P-max}]. 
We will see that investigating normalized quantities such as $\eta/\eta_{C}$ and $P/P_{\Delta T}$ has distinct advantages in comparing linear and nonlinear transport on the same footing; it also allows us to predict nonlinear thermoelectricity based on linear-response quantities. 

\subsection{Linear-response quantities}
\label{sec:linear-response-theory}

Since the formalism of the previous section describes fully nonlinear transport of particle and heat, it readily reproduces the linear response theory by expanding the result regarding small bias and temperature difference. For convenience, we here collect results of linear-response quantities necessary for later analysis, following the notation of Ref.~\cite{Taniguchi20}. 

Within the linear response theory, one can describe thermoelectric transport of electronic contribution by using the formula:
\begin{align}
& h 
\begin{pmatrix} I_{L} \\ \beta J_{L}^{\text{el}} \end{pmatrix}
= \begin{pmatrix}
\mathcal{K}_{0} & \mathcal{K}_{1} \\ \mathcal{K}_{1} & \mathcal{K}_{2}
\end{pmatrix}
\begin{pmatrix} -\Delta \mu \\ k_{B}\Delta T
\end{pmatrix},
\end{align}
by assuming $\Delta T=T_{L}-T_{R}>0$ and $\Delta \mu=\mu_{R}-\mu_{L}>0$ are much smaller than the average temperature $k_{B}\bar{T}=k_{B}(T_{L}+T_{R})/2$. 
Here we have used $\beta=(k_{B}\bar{T})^{-1}$ and the dimensionless Onsager coefficients $\mathcal{K}_{n}$. 
As for coherent transport across a nanostructure, one can express these coefficients in terms of transmission function $\mathcal{T}_{\text{el}}(E)$ as
\begin{align}
  \mathcal{K}_{n} = \beta^{n} \int dE (E-\mu)^{n} \mathcal{T}_{\rm el}(E) \left[ - \frac{\partial f}{\partial E} \right]. 
\end{align}
By these coefficients $\mathcal{K}_{n}$, we can express standard linear-response quantities:
\begin{align}
& G=\frac{e^{2}}{h}\mathcal{K}_{0}, \:\: \kappa_{\text{el}}=\frac{k_{B}^{2}T}{h}\Big( \mathcal{K}_{2}-\frac{\mathcal{K}_{1}^{2}}{\mathcal{K}_{0}} \Big), \:\: S = \frac{k_{B}}{e} \frac{\mathcal{K}_{1}}{\mathcal{K}_{0}}. 
  \label{eq: Linear response quantities}
\end{align}
Therefore, the figure of merit $(ZT)_{\text{el}}$ when ignoring the phonon adverse effect becomes 
\begin{align}
  (ZT)_{\text{el}} = \frac{S^{2}GT}{\kappa_{\text{el}}} = \frac{\mathcal{K}_{1}^{2}}{\mathcal{K}_{0} \mathcal{K}_{2} - \mathcal{K}_{1}^{2}}.
\end{align}
One can also evaluate the output power $P$ by Eq.~\eqref{def:power-efficiency}, and express the stopping bias potential $\Delta \mu_{\text{stop}}$ and maximal power output $P_{\max}$ as
\begin{align}
& \Delta \mu_{\text{stop}}=-e S \Delta T = \frac{\mathcal{K}_{1}}{\mathcal{K}_{0}}k_{B}\Delta T, \\
& P_{\max}=\frac{GS^{2}}{4}(\Delta T)^{2}=P_{\Delta T}\cdot\frac{\mathcal{K}_{1}^{2}}{\mathcal{K}_{0}}.  
\label{eq:P-max}
\end{align}

The total thermal conductance $\kappa$ is the sum of the electron and phonon contributions, $\kappa=\kappa_{\text{el}}+\kappa_{\text{ph}}$, where one can derive phonon conductance $\kappa_{\text{ph}}$ from Eq.~\eqref{eq:J-ph} as
\begin{align}
& \kappa_{\text{ph}}=\int^{\infty}_{0}\frac{dE}{h}E \mathcal{T}_{\text{ph}}(E) \frac{\partial n(E)}{\partial T}. 
\end{align}
It is clear that the presence of $\kappa_{\text{ph}}\gg \kappa_{\text{el}}$ significantly lowers the value of $ZT$ from $(ZT)_{\text{el}}$ by
\begin{align}
& ZT = \frac{S^{2}GT}{\kappa_{\text{el}}+\kappa_{\text{ph}}}
= \frac{(ZT)_{\text{el}}}{1+\kappa_{\text{ph}}/\kappa_{\text{el}}},
\label{eq:ZT-total}
\end{align}
as well as the linear-response efficiency $\eta=\eta_{\text{el}}/(1+\kappa_{\text{ph}}/\kappa_{\text{el}})$.  Thus for electron's quantum coherence to improve thermoelectricity, it is a prerequisite to find a system with thermal conductance satisfying $\kappa_{\text{ph}}\lesssim \kappa_{\text{el}}$. 

We note that when introducing the dimensionless bias voltage $v=\Delta\mu/\Delta\mu_{\text{stop}}$, we can express the output power as $P/P_{\max}=4v(1-v)$ and the electron efficiency as $\eta_{\text{el}}=v(1-v)/[1-v+(ZT)_{\text{el}}^{-1}]$~\cite{Taniguchi20}. As a result, using these dimensionless quantities enables us to estimate the power-efficiency diagram by changing the bias voltage within the linear response theory. 

\section{Search for suitable structures by linear-response quantities}

\label{sec:linear-response}

Our strategy to get higher thermoelectricity in quantum nanostructures is to use electron's destructive quantum interference. However, to make such an effect conspicuous, we should suppress phonon conductance $\kappa_{\text{ph}}$ smaller than $\kappa_{\text{el}}$ [see Eq.~\eqref{eq:ZT-total}]. 
%
Temperature increase in $\kappa_{\text{ph}}$ is usually much faster than that in $\kappa_{\text{el}}$. We have observed that though it highly depends on the location of the electrochemical potential, a typical value of electron's thermal conductance $\kappa_{\text{el}}$ amounts to $\SI{e-12}{WK^{-1}}$ or less around \SI{10}{K} for a quantum dot considered here [see Eq.~\eqref{eq: Linear response quantities}]. This implies that we usually have difficulty in finding a temperature range suitable to suppress $\kappa_{\text{ph}}$. 
Nevertheless, we will show that modifying bending at the junction can create a situation $\kappa_{\text{ph}}\lesssim \kappa_{\text{el}}$, where we will further vary local gate voltage $\epsilon_{g}$ to achieve higher $ZT$ and thermal efficiency. 

\subsection{Phonon thermal conductance}

\begin{figure}
  \centering
  \subfloat[\label{fig:phonon-conductance-N15}]{
    \includegraphics[width=0.48\linewidth,trim=10 10 10 100]{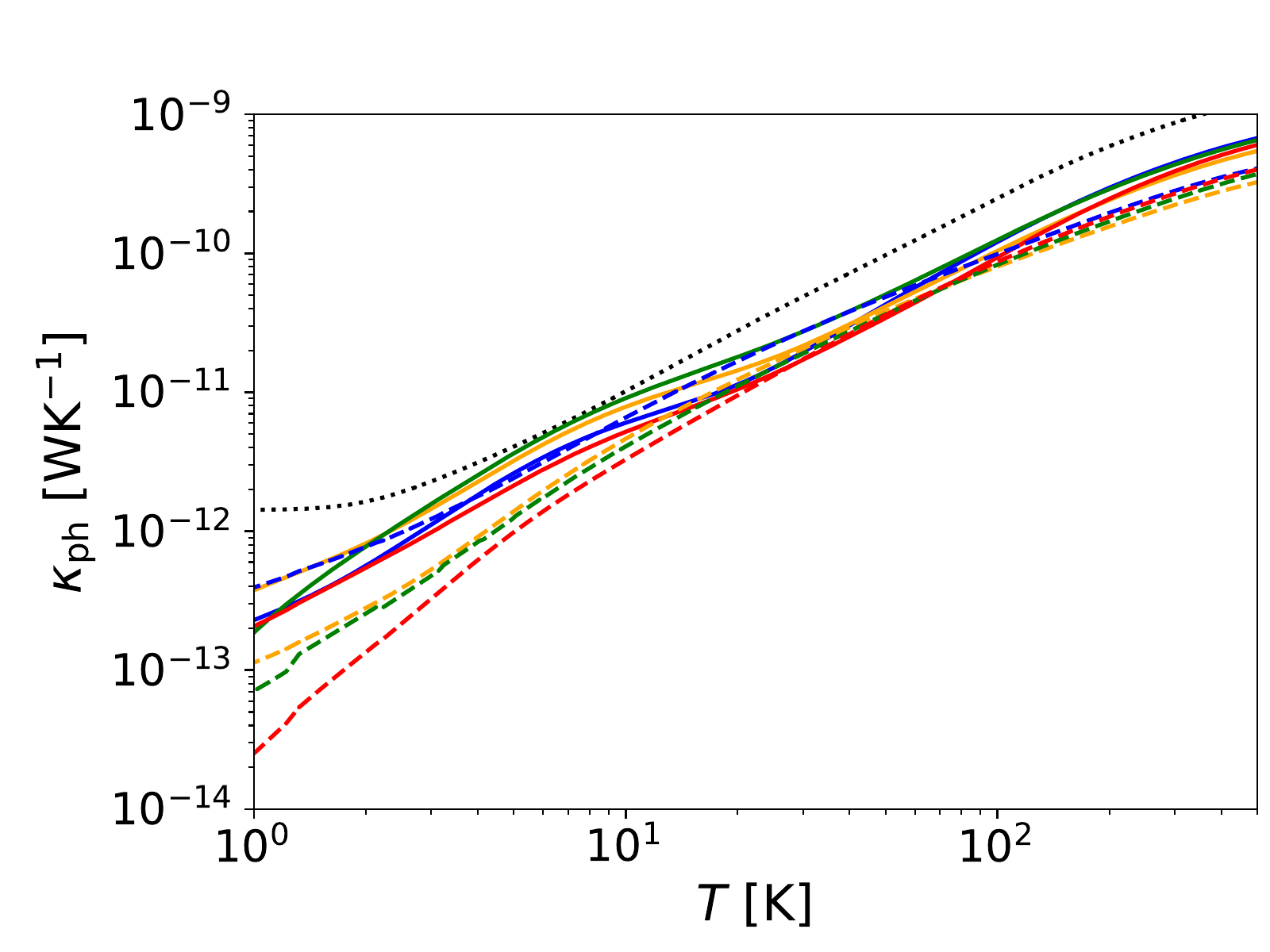}
  }
  \subfloat[\label{fig:phonon-conductance-N13}]{
    \includegraphics[width=0.48\linewidth,trim=10 10 10 100]{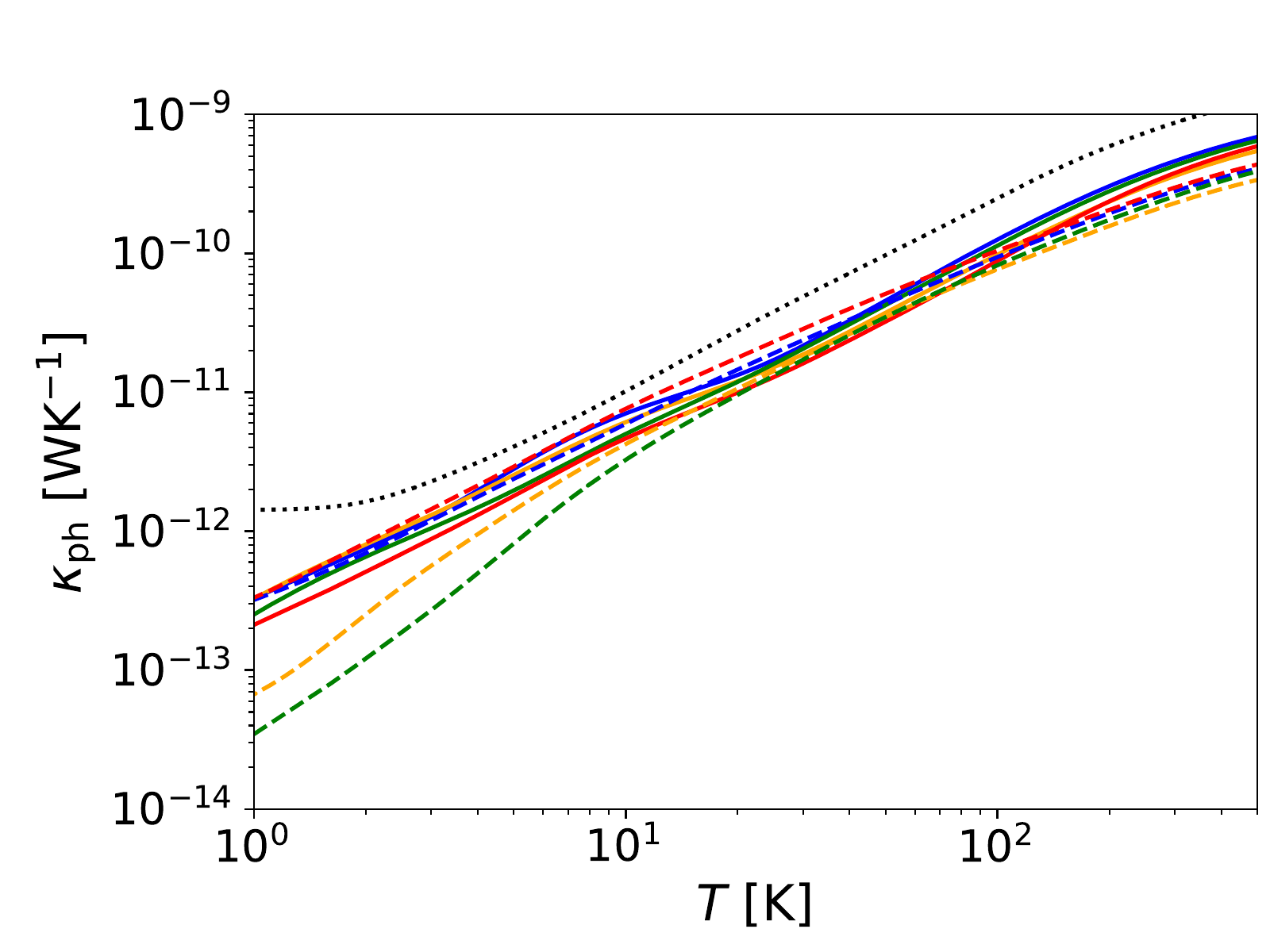}
  }
  \\ 
  \subfloat[\label{fig:phonon-conductance-N11}]{
    \includegraphics[width=0.48\linewidth,trim=10 10 10 50]{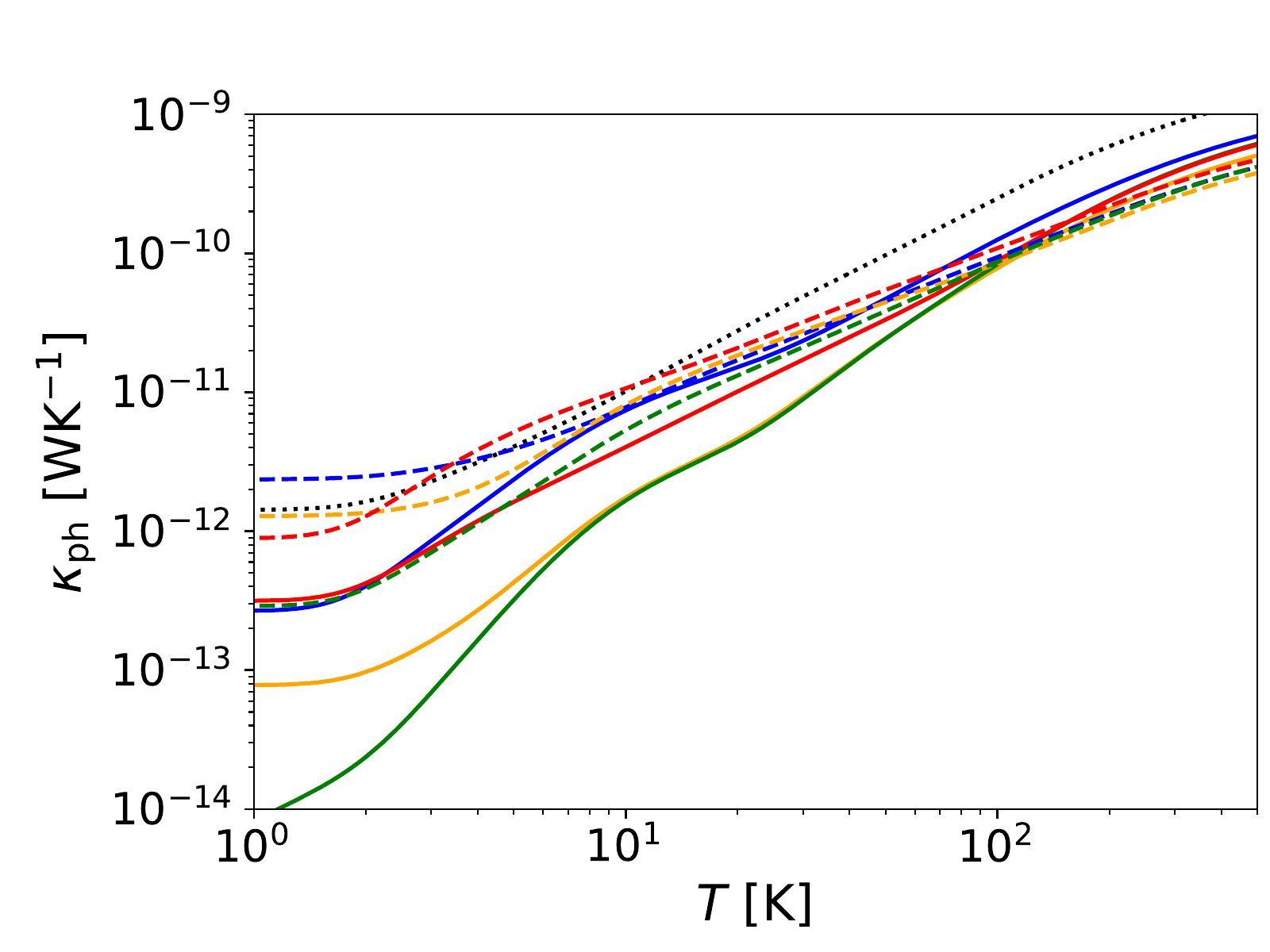}
  }
  \subfloat[\label{fig:phonon-conductance-N09}]{
    \includegraphics[width=0.48\linewidth,trim=10 10 10 50]{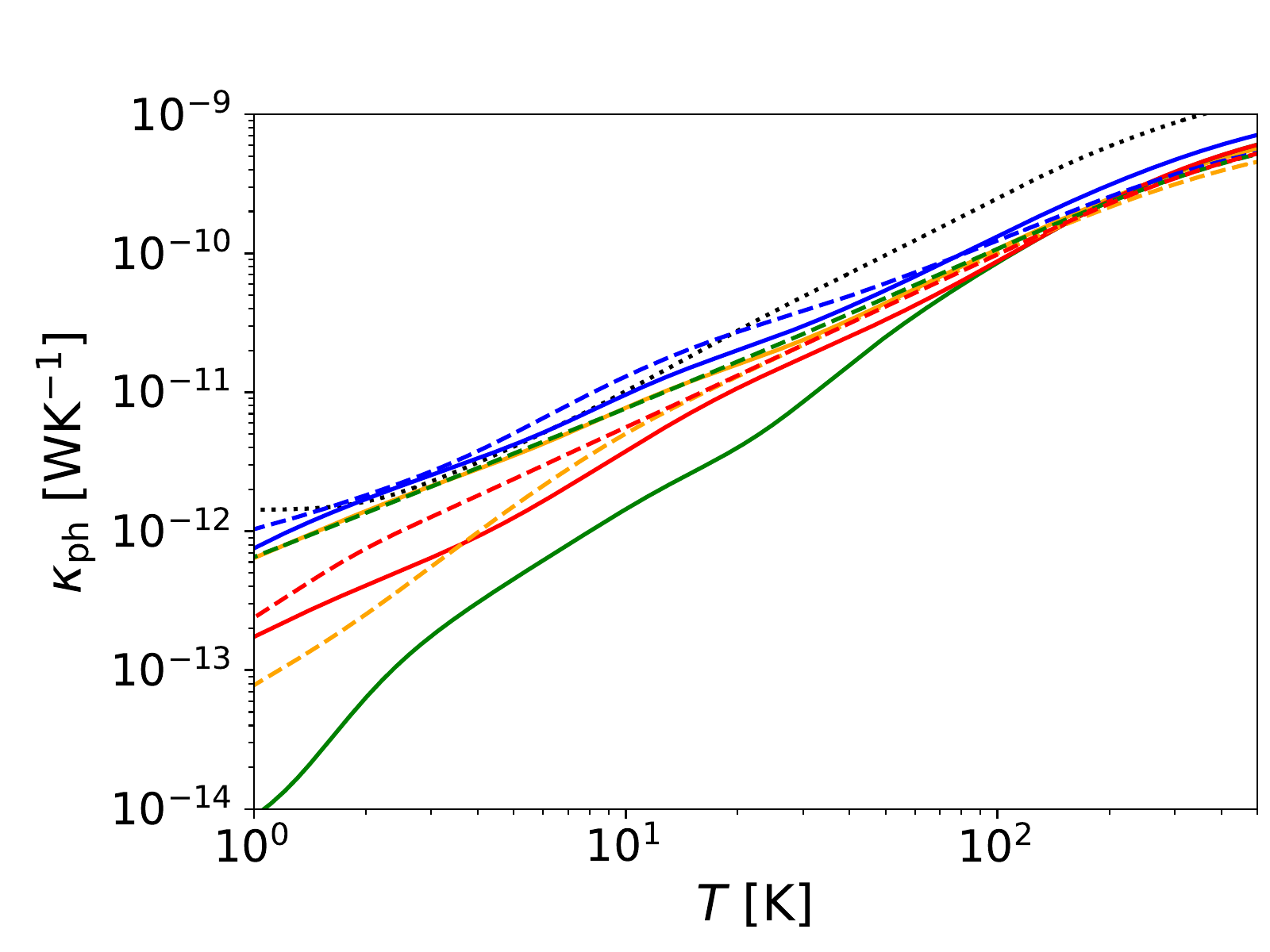}
  }
  \caption{Temperature dependence of phonon thermal conductance of a graphene rhombus dot (solid line) and a graphene ring (dashed line) with four types of bending angles of junctions: simple-attached (blue), soft-bent (orange), hard-bent (red) and double-bent (green). Black dotted line refers to the result of a simple nanoribbon. The size of a rhombus is (a) $N=15$, (b) $N=13$, (c) $N=11$ and (d) $N=9$. 
}
\label{fig:phonon-conductance}
\end{figure}

Let us start by examining how various modifications of a nanostructure affect phonon thermal conductance. Figure~\ref{fig:phonon-conductance} shows the temperature dependence of phonon thermal conductance for a rhombus dot (solid lines) and a rhombus ring (dashed lines). Different bending configurations of the junction (as in Fig.~\ref{fig:four-bending-angles}) are shown in different colors: simple-attached (blue), soft-bent (orange), hard-bent (red), and double-bent (green). For reference, we also include the result of a straight nanoribbon with the width $w_{0}=\SI{0.738}{nm}$ (black dotted line).  Besides, we vary the size of a rhombus itself: (a) $N=15$, (b) $N=13$, (c) $N=11$ and (d) $N=9$, as is defined in Fig.~\ref{fig:rhombus-size}. Depending on the relative size of a structure to the lead width, we will see different effects of the bending on phonon conductance. 

At high temperatures $T\gtrsim \SI{100}{K}$, phonon conductance in all the configurations reaches $\SI{e-10}{WK^{-1}}$. The value greatly exceeds a typical value of $\kappa_{\text{el}}$, though it is smaller than the phone conductance of a stright nanoribbon. For larger sizes ($N=15$ and $N=13$ in Figs.~\ref{fig:phonon-conductance-N15} and \ref{fig:phonon-conductance-N13}), phonon conductance of a rhombus ring is smaller than that of a rhombus dot, reflecting the increased scattering by a puncture inside the rhombus. For smaller sizes ($N=11$ and $N=9$ in Figs.~\ref{fig:phonon-conductance-N11} and \ref{fig:phonon-conductance-N09}), phonon conductance seems independent of whether a rhombus dot or ring. This suggests that phonons in these systems are mainly scattered by the bending, not by the puncture inside the rhombus. With $\kappa_{\text{ph}}\gg \kappa_{\text{el}}$ in this temperature range, we find it challenging to enhance thermoelectricity by controlling the electron's coherency. 

The situation differs at low temperatures $T\lesssim \SI{10}{K}$,
especially for smaller rhombus ($N=11$ and $N=9$ in
Figs.~\ref{fig:phonon-conductance-N11} and
\ref{fig:phonon-conductance-N09}). Phonon conductance depends highly
on the bending angle at the junction, compared to larger rhombuses
($N=15$ and $N=13$).  We observe that the phonon conductance of the
double-bent rhombus dot is significantly small, reaching an order of
$\SI{e-13}{WK^{-1}}$ or less. This is one order of magnitude smaller
than that of the simple-attached rhombus dot.  These results are
consistent with Refs.~\cite{Xu10,Li14,Savin10}, which attributed the
reduction of $\kappa_{\text{ph}}$ to phonon scattering due to
interface mismatching and rough-edge effects. We emphasize the value
of $\kappa_{\text{ph}}$ of the double-bent rhombus dot is comparable
or smaller than $\kappa_{\text{el}}$ at $T\lesssim
\SI{10}{K}$. Accordingly, using a double-bent graphene rhombus is a
viable strategy to suppress phonon transport. In the next section, we
will exploit quantum coherence to get a better thermoelectricity.

\subsection{Linear-response thermoelectricity}
\label{sec:linear-thermoelectricity}

Having identified the structure suitable for suppressing phonon transport, we will now demonstrate how to improve its thermoelectricity using the local gate voltage $\epsilon_{g}$. Based on the result of phonon conductance in the previous section, we choose to operate the double-bent graphene rhombus dot of $N=11$ at temperature $T=\SI{4}{K}$, whose phonon conductance $\kappa_{\text{ph}}$ is low (Fig.~\ref{fig:phonon-conductance-N11}).  To clarify how different bending angles affect $ZT$, we compare it with the simple-attached rhombus dot of the same size.

\begin{figure}{tb}
  \centering
  \subfloat[\label{fig: ZT_el, straight}]{
    \includegraphics[width=0.49\linewidth]{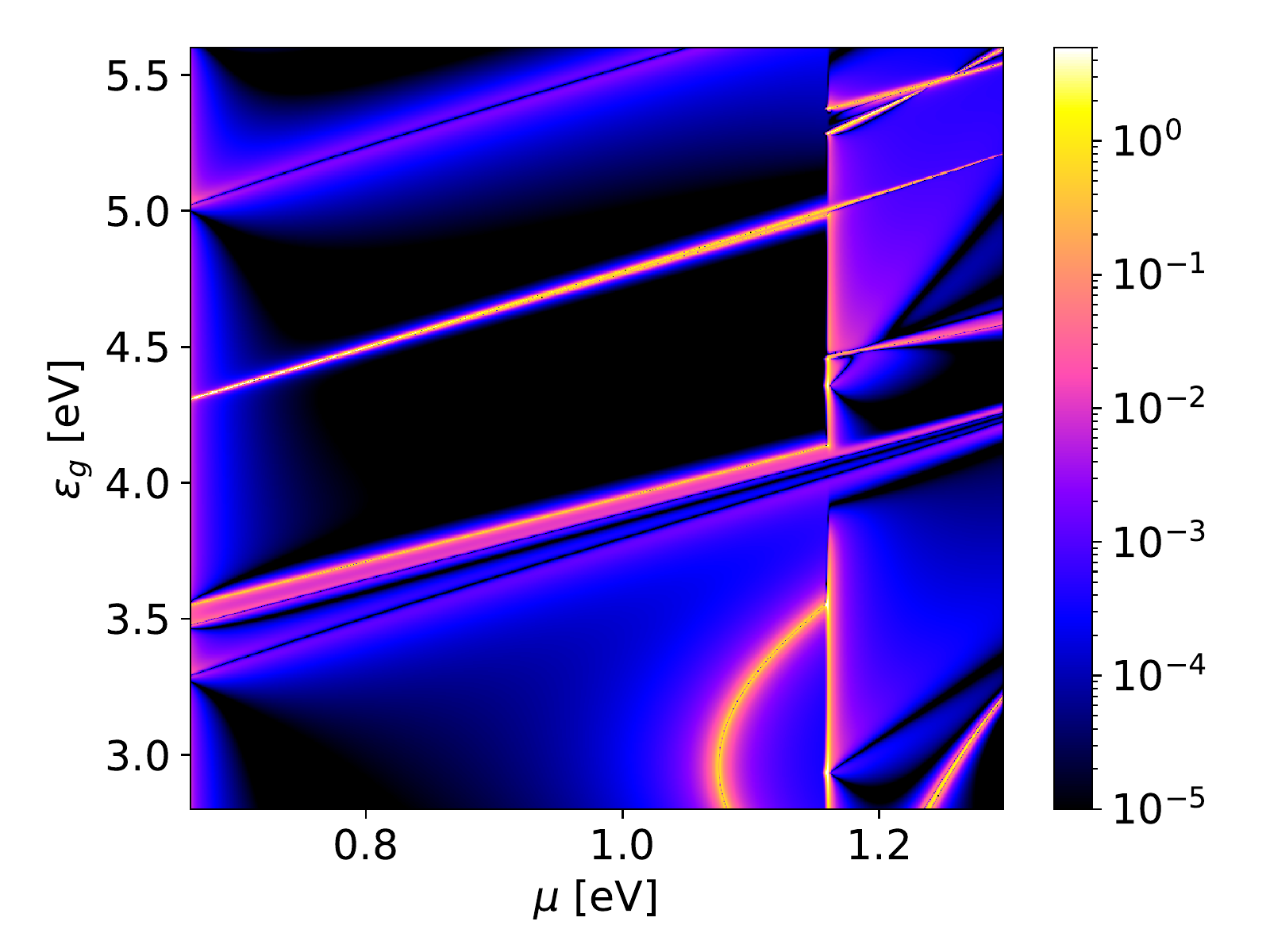}
  }
  \subfloat[\label{fig: ZT, straight}]{
    \includegraphics[width=0.49\linewidth]{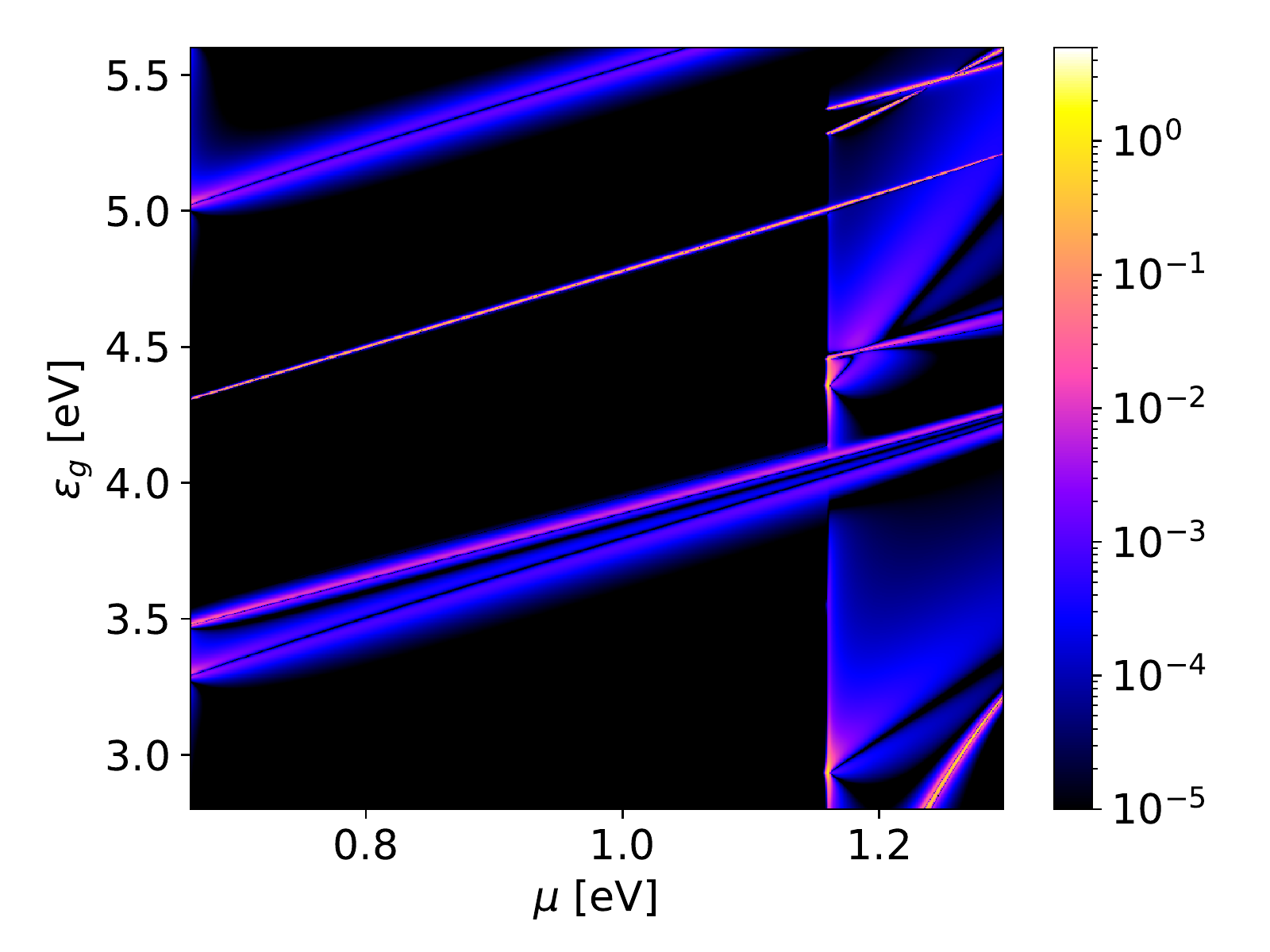}
  }
  \caption{The figure of merit of the simple-attached rhombus dot as a function of $\mu$ and $\epsilon_{g}$. (a) $(ZT)_{\text{el}}$, neglecting phonon transport and (b) $ZT$ including phonon contribution.
  \label{fig:ZT-for-simple-attached}}
\end{figure}
\begin{figure}
  \centering
  \subfloat[\label{fig: ZT_el, double-bend}]{
    \includegraphics[width=0.49\linewidth]{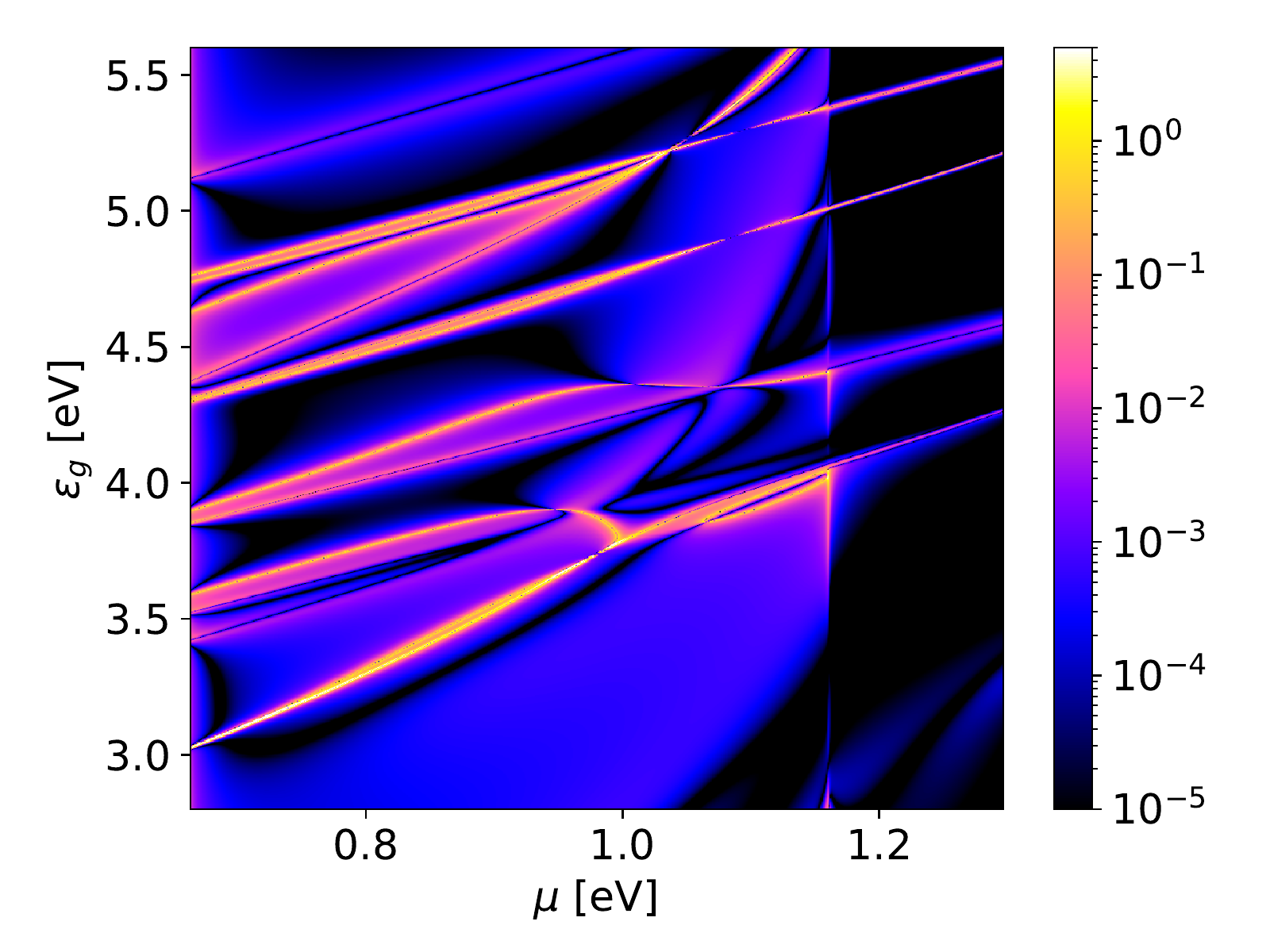}
  }
  \subfloat[\label{fig: ZT, double-bend}]{
    \includegraphics[width=0.49\linewidth]{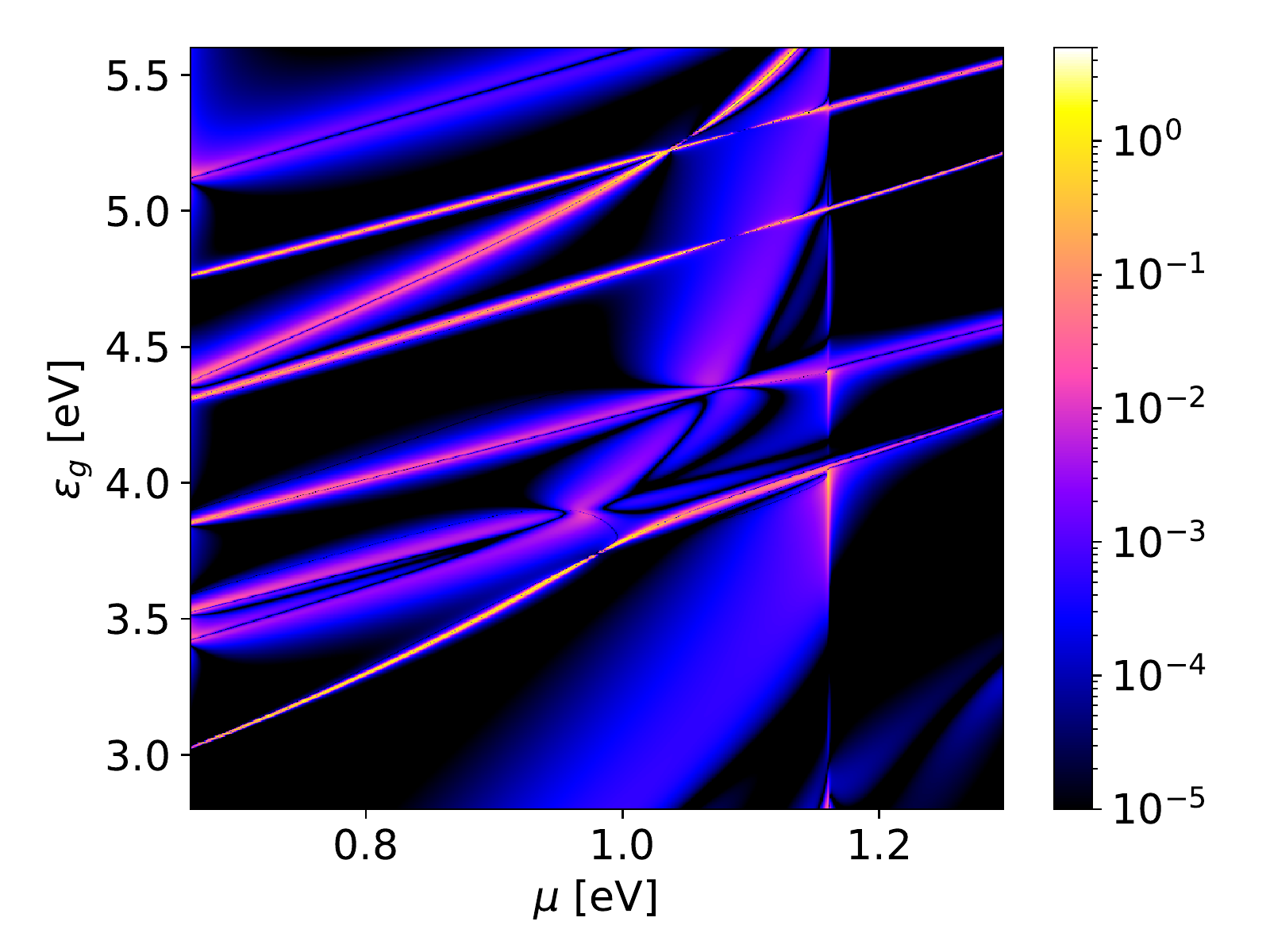}
  }
  \caption{The figure of merit of the double-bent rhombus dot as a function of $\mu$ and $\epsilon_{g}$. (a) $(ZT)_{\text{el}}$ without phonon transport and (b) $ZT$ including phonon contribution.}
\label{fig:ZT-for-double-bent}
\end{figure}

Figures \ref{fig:ZT-for-simple-attached} and \ref{fig:ZT-for-double-bent} show how the figure of merit depends on the electrochemical potential $\mu$ and the local gate voltage $\epsilon_{g}$ for the simple-attached and double-bent rhombus dots. In each figure, we compare (a) the electron contribution $(ZT)_{\text{el}}$ with (b) the total contribution $ZT$ that includes phonon transport. The result of the simple-attached dot (Fig.~\ref{fig:ZT-for-simple-attached}) shows that though the value of $(ZT)_{\text{el}}$ amounts to well above $1$ (even reaching above $5$), large phonon conductance $\kappa_{\text{ph}}$ considerably reduces its value. For instance, high $(ZT)_{\text{el}}$ around the region $(\mu,\epsilon_{g})\sim (\SI{1.1}{eV}, \SI{3.0}{eV})$ or $(\SI{1.0}{eV}, \SI{4.4}{eV})$ does not lead to a high value of $ZT$. The maximum of $ZT$ in the parameter range of Fig.~\ref{fig:ZT-for-simple-attached} is $1.1$. In contrast, the double-bent rhombus dot is more robust against the phonon deterioration effect due to a smaller value of $\kappa_{\text{ph}}$. Indeed, $ZT$ reaches as much as $3.1$ by adjusting the local gate voltage $\epsilon_{g}\approx \SI{5.40}{eV}$.
\begin{figure}
  \centering
  \subfloat[\label{fig:ZT-and-transmission-for-simple-attached}]{
    \includegraphics[width=0.5\linewidth,trim=80 10 50 50]{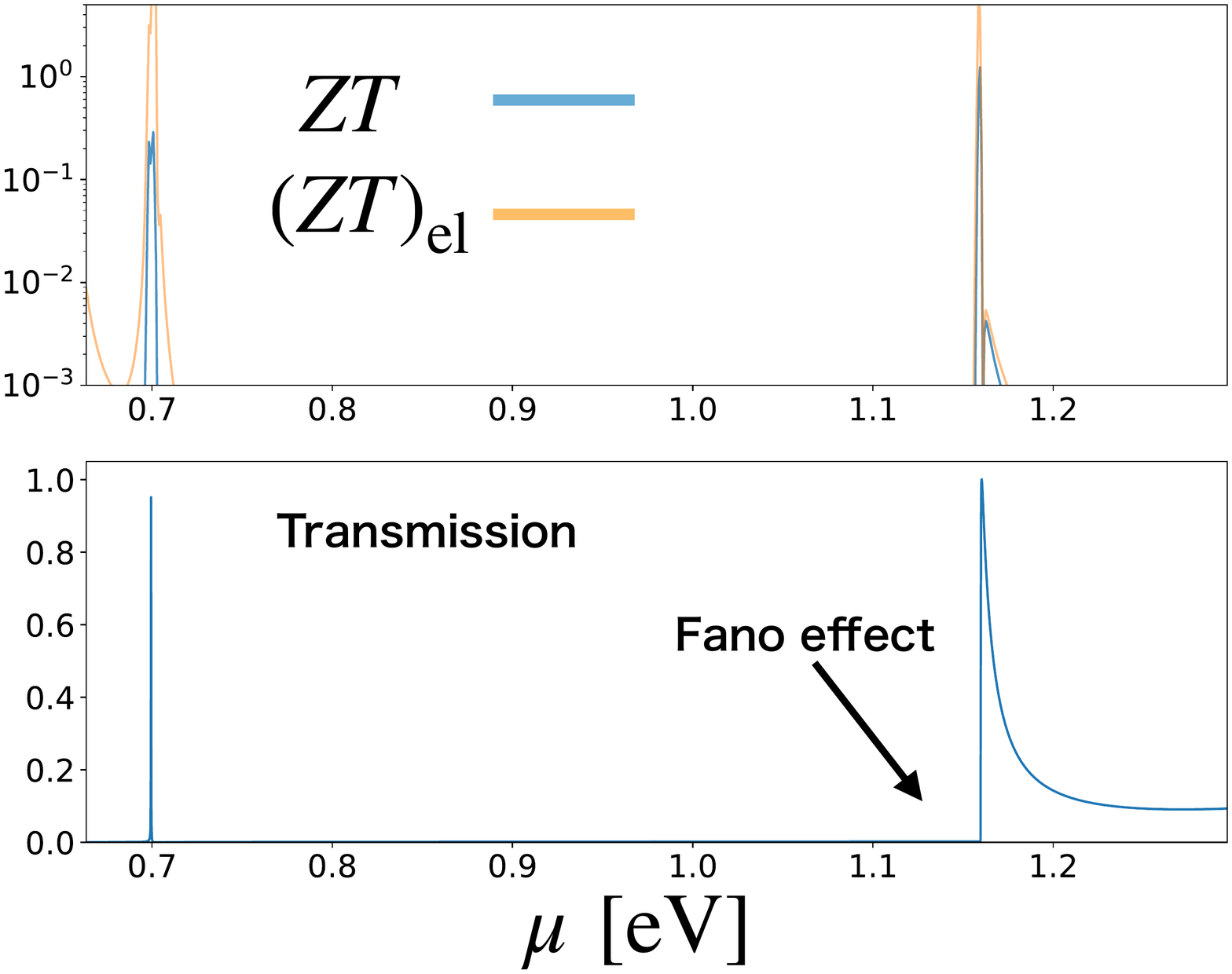}
  }
  \subfloat[\label{fig:ZT-and-transmission-for-double-bent}]{
    \includegraphics[width=0.5\linewidth,trim=80 10 50 50]{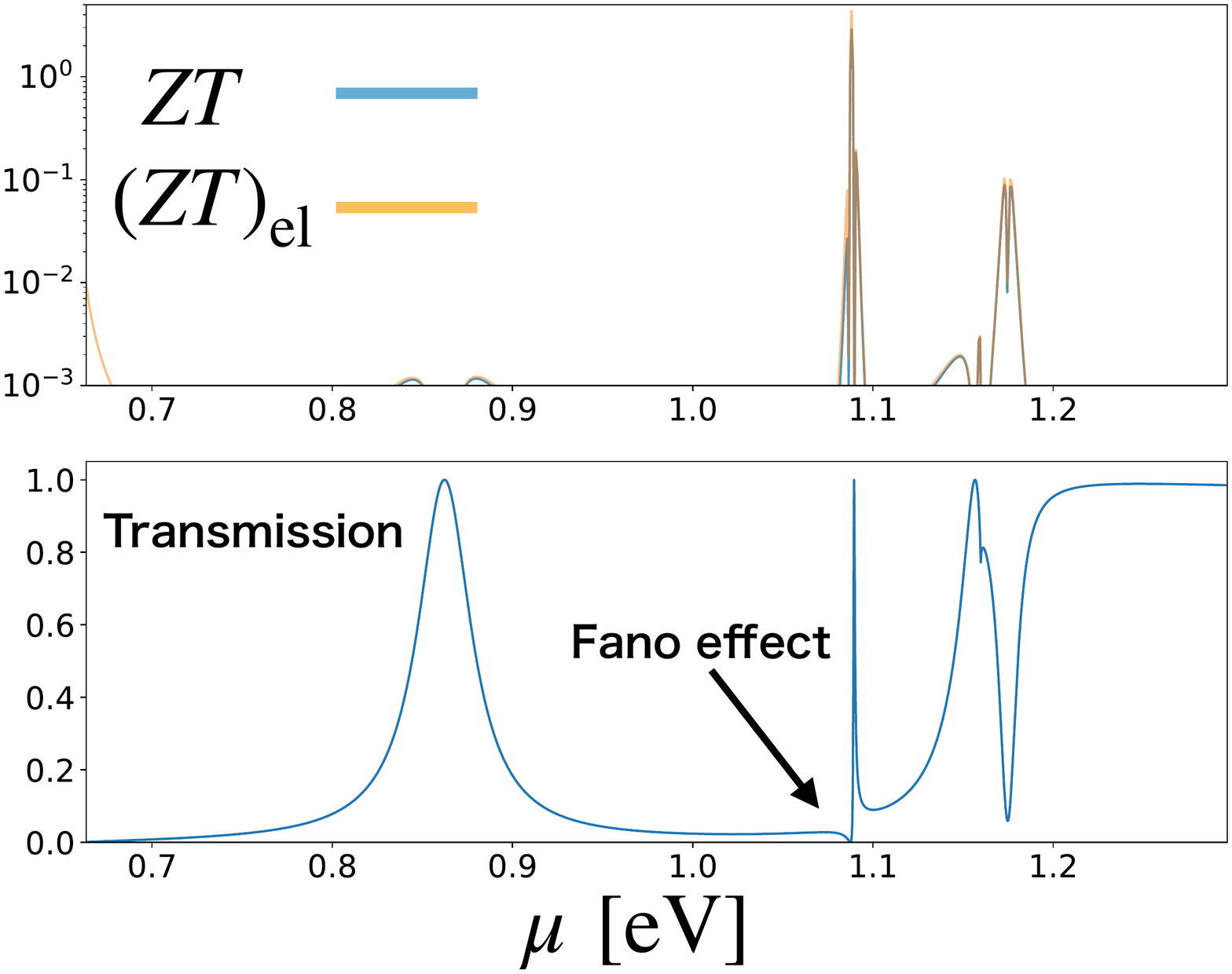}
  }
  \caption{Comparison of the figures of merit, $ZT$ (blue) and $(ZT)_{\text{el}}$ (orange), with the transmission function $\mathcal{T}(\mu)$ (below), as a function of chemical potential $\mu$. (a) the simple-attached rhombus dot at $\epsilon_{g}=\SI{4.36}{eV}$ and (b) the double-bent rhombus dot at $\epsilon_{g}=\SI{5.40}{eV}$.
\label{fig:ZT-and-transmission}}
\end{figure}

It is worthwhile to inspect what causes high values of $ZT$ in the presence of phonon transport. We observe that the Fano-type transmission is responsible for it. In Fig.~\ref{fig:ZT-and-transmission}, we compare the figure of merit with the transmission spectrum as a function of the chemical potential $\mu$ at the fixed local gate voltage that achieves the highest value of $ZT$: (a) $\epsilon_{g}=\SI{4.36}{eV}$ for the simple-attached dot and (b) $\epsilon_{g}=\SI{5.40}{eV}$ for the double-bent dot. We see a Fano-type asymmetric resonance occur at $\mu\approx \SI{1.15}{eV}$ in Fig.~\ref{fig:ZT-and-transmission-for-simple-attached} or $\mu\approx \SI{1.09}{eV}$ in Fig.~\ref{fig:ZT-and-transmission-for-double-bent}, as well as a Breit-Wigner-type symmetric one at $\mu\approx \SI{0.7}{eV}$ in Fig.~\ref{fig:ZT-and-transmission-for-simple-attached} or $\mu\approx\SI{0.86}{eV}$ in Fig.~\ref{fig:ZT-and-transmission-for-double-bent}. 
Clearly, both types of resonances can produce high $(ZT)_{\text{el}}$ in the 
absence of the phonon degradation effect. 
However, Fano-type resonances provide higher $ZT$, making them much more robust against phonon transport than Breit-Wigner ones.
This is seen from the results of the simple-attached dot (Fig.~\ref{fig:ZT-and-transmission-for-simple-attached}). A very narrow Breit-Wigner resonance at $\mu\approx \SI{0.7}{eV}$ produces a high $(ZT)_{\text{el}}$, but the total $ZT$ gets suppressed by more than one order of magnitude from $(ZT)_{\text{el}}$ due to the phonon degradation effect.  In contrast, the suppression of $ZT$ at $\mu\approx \SI{1.15}{eV}$ is not so drastic, though the relatively large value of $\kappa_{\text{ph}}$ in the simple-attached dot makes $ZT$ smaller than one. 
In the double-bent dot (Fig.~\ref{fig:ZT-and-transmission-for-double-bent}), which has a reduced value of $\kappa_{\text{ph}}$, the degradation of $ZT$ due to phonons becomes less striking at the Fano resonance ($\mu\approx \SI{1.09}{eV}$).
The Breit-Wigner resonance at $\mu\approx \SI{0.86}{eV}$ does not produce high $ZT$ or $(ZT)_{\text{el}}$ because the resonance width is too large. 
 It shows that suppressing the phonon transport is an effective way to utilize the enhanced thermoelectricity due to electron's quantum coherence and Fano resonances. 
%
%

\begin{figure}
  \centering
  \subfloat[\label{fig: eta vs power, linear, straight}]{
    \includegraphics[width=0.48\linewidth]{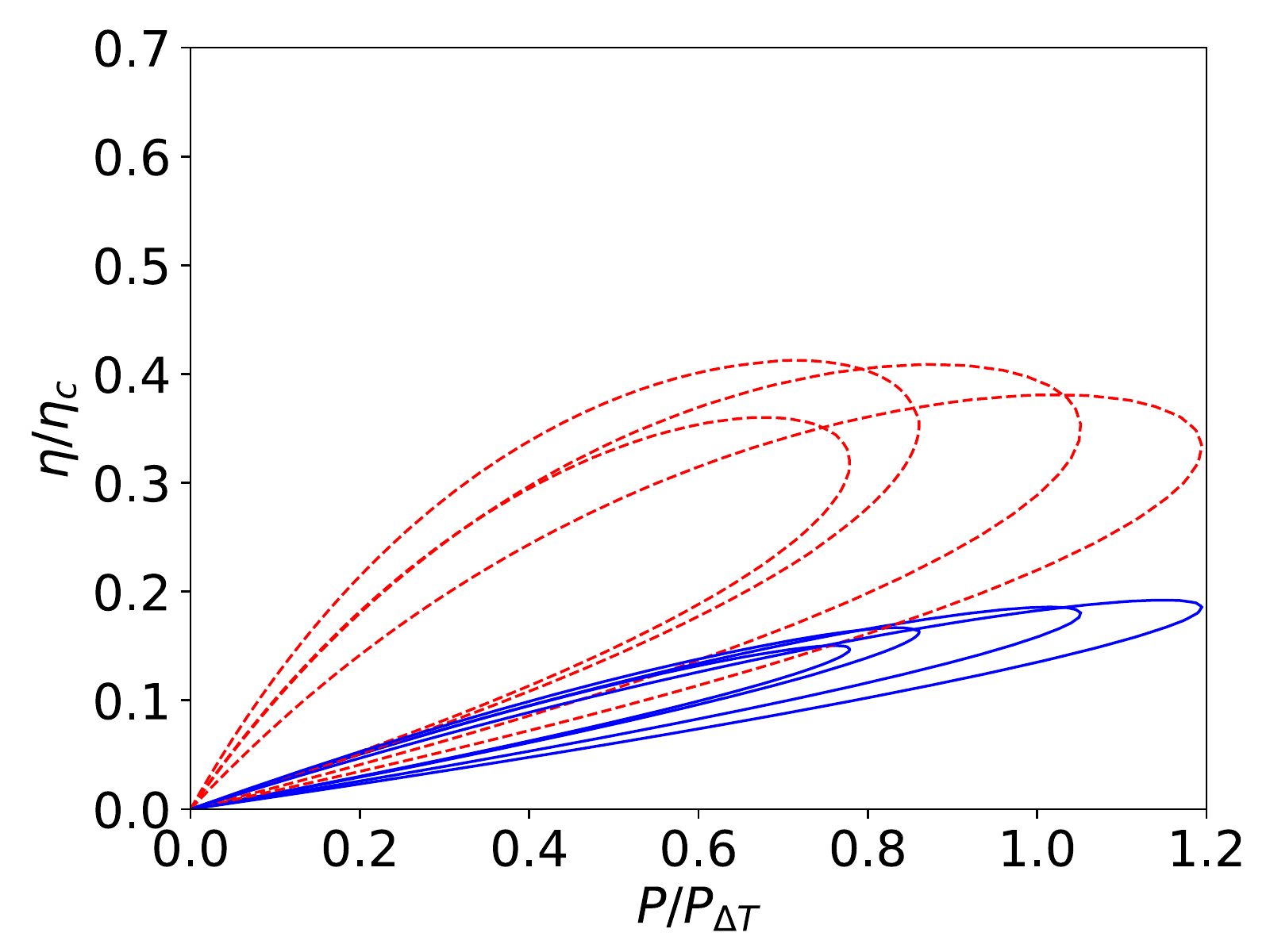}
  }
  \subfloat[\label{fig: eta vs power, linear, double-bend}]{
    \includegraphics[width=0.48\linewidth]{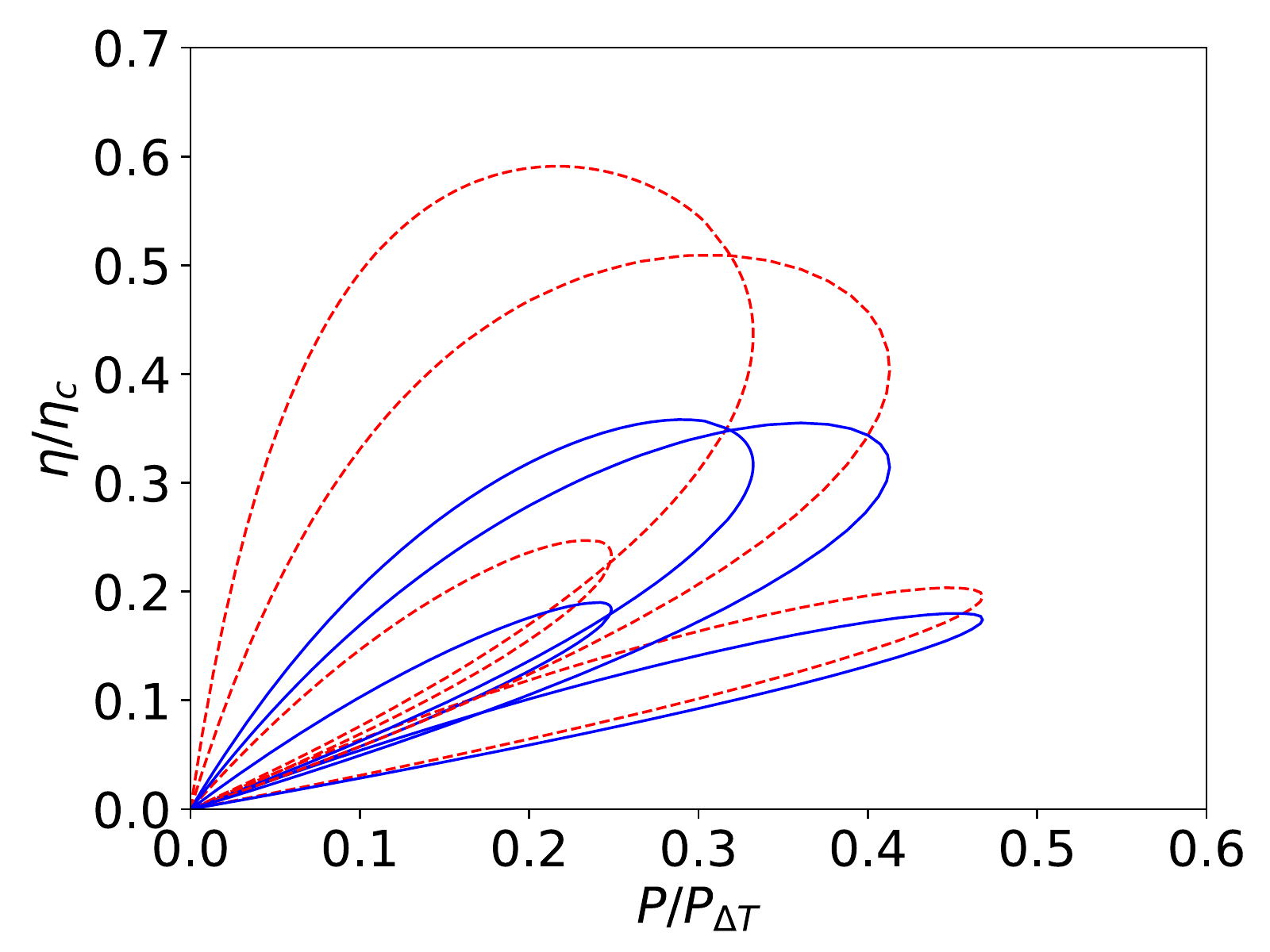}
  }
\caption{Power-efficiency diagram within the linear-response theory for (a) the simple-attached rhombus dot at $\epsilon_{g}=\SI{4.36}{eV}$ and (b) the double-bent rhombus dot at $\epsilon_{g}=\SI{5.40}{eV}$. Each blue line corresponds to the evolution of the power-efficiency $(P,\eta)$ by changing the bias voltage at a fixed $\mu$. The result is compared with the evolution of $(P,\eta_{\text{el}})$ ignoring phonon transport (red dashed line). Efficiency and output power are normalized by $\eta_{c}$ and $P_{\Delta T}$ respectively.
}
\label{fig:linear-power-efficiency-diagram}
\end{figure}

To realize a heat engine, we need to attain high output power besides high efficiency. One can assess such performance by examining the power-efficiency diagram $(P,\eta)$. In Fig.~\ref{fig:linear-power-efficiency-diagram}, we draw the power-efficiency diagram within the linear response theory, (a) for the simple-attached rhombus at $\epsilon_{g}=\SI{4.36}{eV}$, and (2) for the double-bent rhombus dot at $\epsilon_{g}=\SI{5.40}{eV}$. Each line corresponds to the evolution by changing the bias voltage at a fixed chemical potential (from $\SI{0.66}{eV}$ to $\SI{1.3}{eV}$). Here, we have normalized the efficiency $\eta$ by the Carnot efficiency and the output power $P$ by $P_{\Delta T}$. Such normalization will later allow us to compare the linear-response result directly with the performance in the fully nonlinear regime. Figure~\ref{fig:linear-power-efficiency-diagram} shows that high output power and high efficiency are well-balanced. Compared with the simple-attached dot, we see that the output power of the double-bent rhombus dot gets lower, though its efficiency is higher. 

Summarizing the linear-response thermoelectricity, we find the double-bent graphene rhombus dot is a promising nanostructure for achieving nanoscale heat engines.  Modifying the bending angle and adjustment of the local gate voltage significantly enhance the thermal efficiency while retaining high output power. It suggests that utilizing the Fano-type asymmetric resonance is a viable option for achieving high thermoelectric performance. 

\section{Nonlinear thermoelectricity}

\label{sec:nonlinear-response} 

Next, we will examine the nonlinear thermoelectric performance of the graphene double-bent rhombus dot ($N=11$). We choose its local gate voltage to be $\epsilon_{g}=\SI{5.40}{eV}$, which has exhibited the highest value of $ZT$ in Sec.~\ref{sec:linear-thermoelectricity}. We focus on nonlinear efficiency and output power in two situations: (1) at a fixed average temperature $\bar{T}=\SI{4}{K}$ by changing thermal bias and (2) at fixed Carnot efficiencies ($\eta_{c}=1/3$ and $2/3$) by changing the average temperature.  To compare them with the linear-response result Fig.~\ref{fig: eta vs power, linear, double-bend}, we normalize the thermal efficiency $\eta$ by $\eta_{c}$ and the output power $P$ by $P_{\Delta T}$.

\subsection{Nonlinear effect at a fixed average temperature}

Figure \ref{fig: eta vs power, T_bar is const, double-bend} shows the efficiency-power diagram for the double-bent graphene rhombus dot at the average temperature $\bar{T}=\SI{4}{K}$, with increasing nonlinearity: (a)~$(T_{L},T_{R}) = (\SI{4.8}{K}, \SI{3.2}{K})$ with $\eta_{c}=1/3$ and (b)~$(T_{L},T_{R})=(\SI{6.0}{K}, \SI{2.0}{K})$ with $\eta_{c}=2/3$. 
Although the efficiency gets suppressed by finite phonon thermal transport (compare blue lines and red dashed lines), maximum efficiency reaches $0.395\eta_{c}$ at $\eta_{c}=1/3$ or $0.426\eta_{c}$ at $\eta_{c}=2/3$. We note that the normalized thermal efficiency $\eta/\eta_{c}$ increases slightly with increasing nonlinearity, which confirms the robustness of enhanced thermal efficiency due to the Fano-type resonance. In contrast, the normalized output power $P/P_{\Delta T}$ stays almost independent of nonlinearity. 

\begin{figure}
  \centering
  \subfloat[\label{fig: eta vs power, (5, 3), double-bend}]{
    \includegraphics[width=0.48\linewidth]{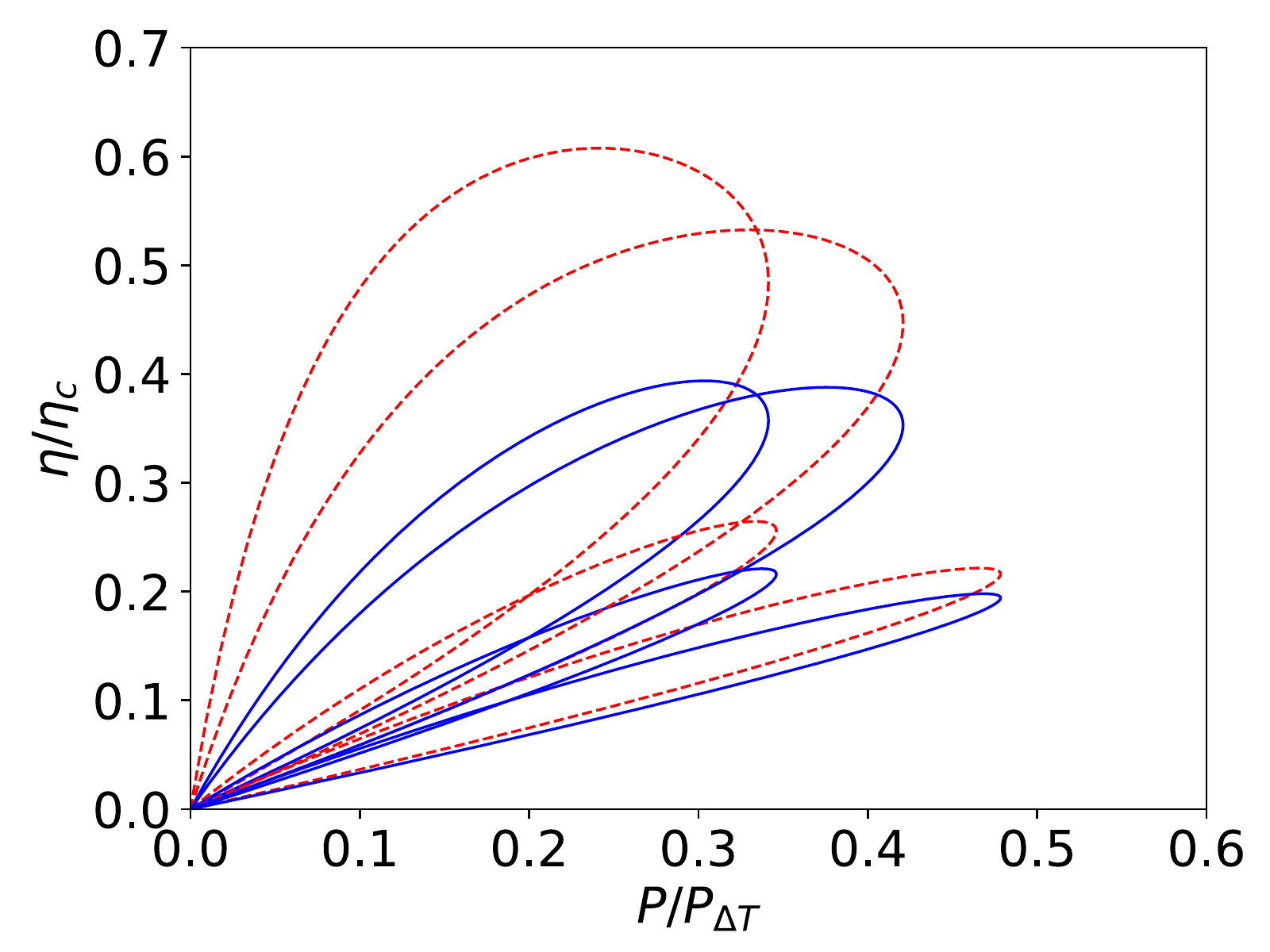}
  }
  \subfloat[\label{fig: eta vs power, (6, 2), double-bend}]{
    \includegraphics[width=0.48\linewidth]{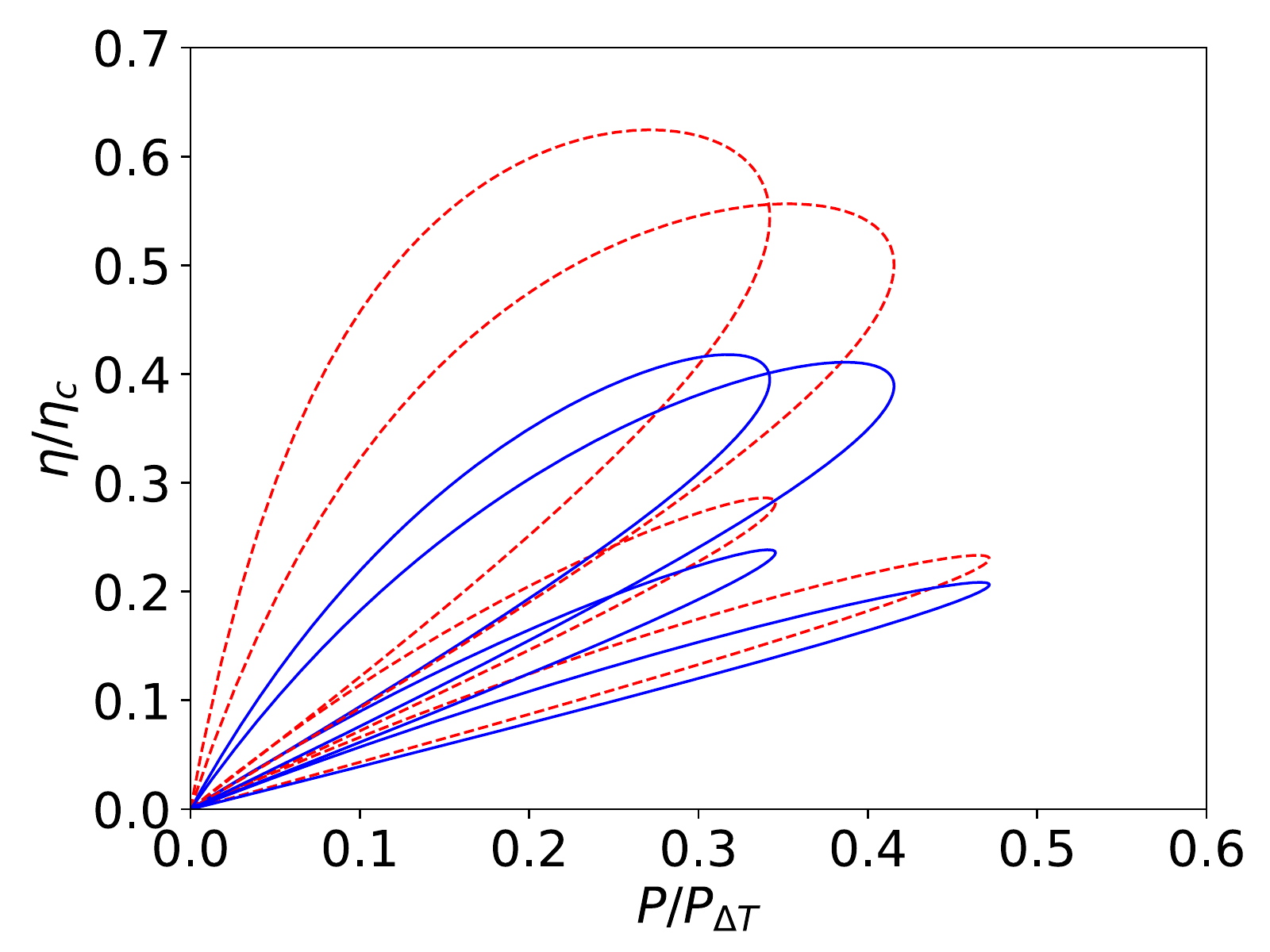}
  }
  \caption{Power-efficiency diagram at the average temperature $\bar{T}=\SI{4}{K}$ for (a) $(T_{L},T_{R})=(\SI{4.8}{K},\SI{3.2}{K})$ with $\eta_{c}=1/3$, and (b) $(T_{L},T_{R})=(\SI{6.0}{K},\SI{2.0}{K})$ with $\eta_{c}=2/3$. Other parameters and conventions are the same with Fig.~\ref{fig: eta vs power, linear, double-bend}.
}
  \label{fig: eta vs power, T_bar is const, double-bend}
\end{figure}

\subsection{Temperature effect at a fixed nonlinearity}

\begin{figure}
  \centering
  \subfloat[\label{fig: eta vs power, (12, 4), double-bend}]{
    \includegraphics[width=0.48\linewidth]{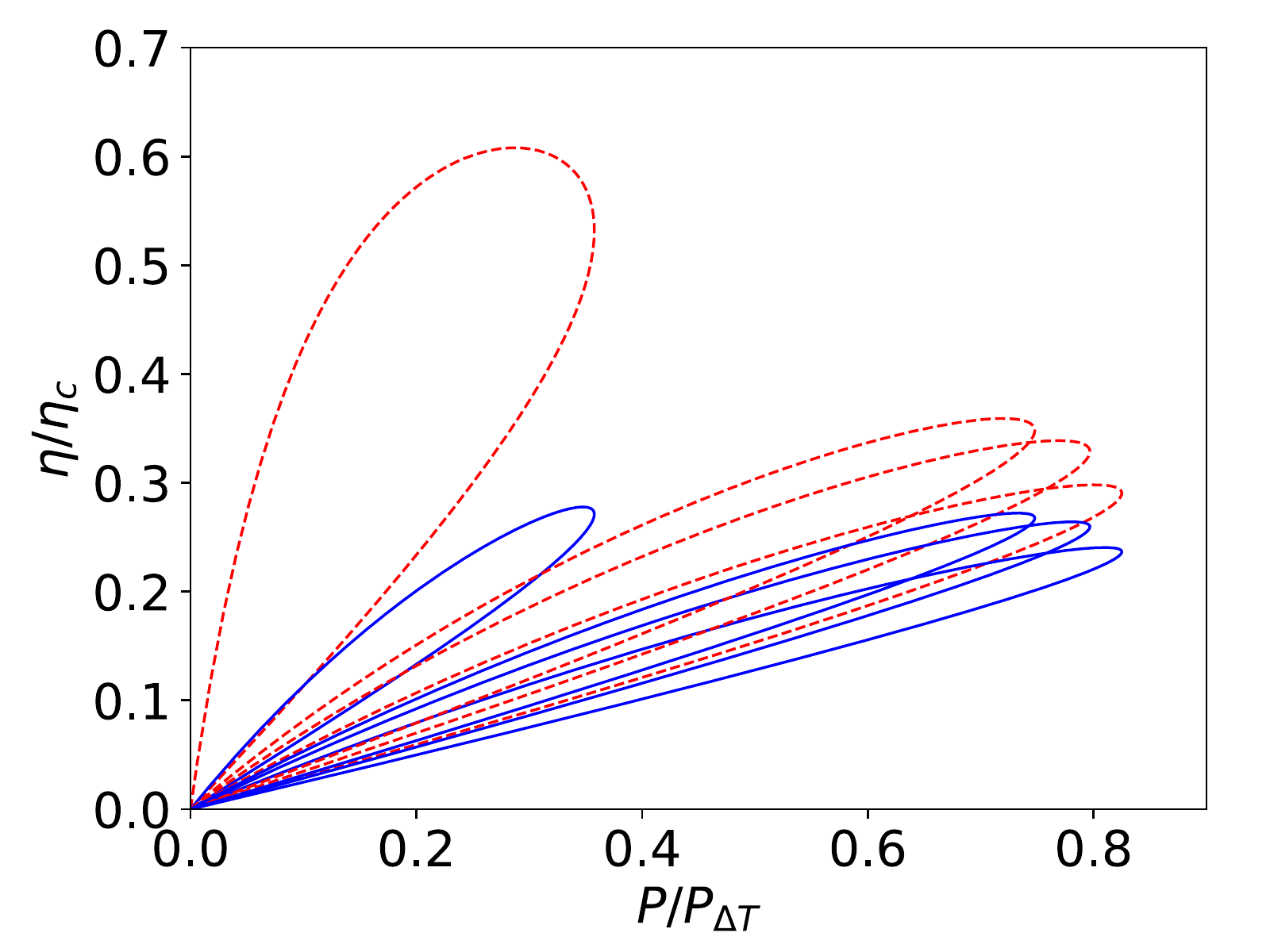}
  }
  \subfloat[\label{fig: eta vs power, (18, 6), double-bend}]{
    \includegraphics[width=0.48\linewidth]{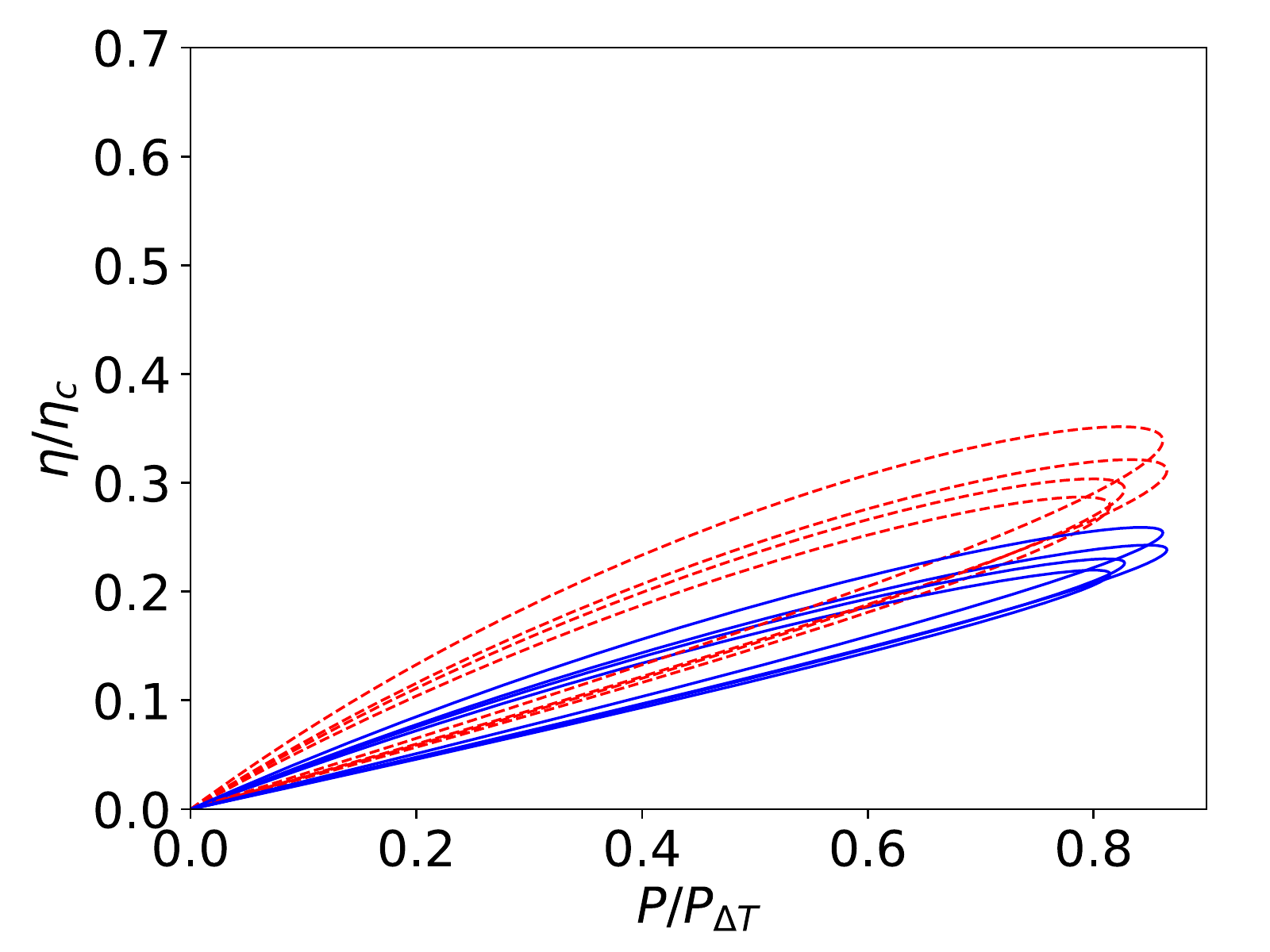}
  }
  \caption{Power-efficiency diagram at the fixed Carnot efficiency $\eta_{c}=2/3$ for (a) $(T_{L},T_{R})=(\SI{12}{K},\SI{4}{K})$ with $\bar{T}=\SI{8}{K}$ and (b) $(T_{L},T_{R})=(\SI{18}{K},\SI{6}{K})$ with $\bar{T}=\SI{12}{K}$. Other parameters and conventions are the same with Fig.~\ref{fig: eta vs power, linear, double-bend}.}
\label{fig:power-efficiency-at-eta=2/3}
\end{figure}

We will now explore the role of the average temperature at a fixed Carnot efficiency. Figure~\ref{fig:power-efficiency-at-eta=2/3} shows the power-efficiency diagrams for the Carnot efficiency $\eta_{c}=2/3$ by increasing the average temperature $\bar{T}$: (a) $(T_{L},T_{R})=(\SI{12}{K}, \SI{4}{K})$ with $\bar{T}=\SI{8}{K}$, and (b) $(T_{L},T_{R})=(\SI{18}{K}, \SI{6}{K})$ with $\bar{T}=\SI{12}{K}$. Comparing these results with Fig.~\ref{fig: eta vs power, (6, 2), double-bend} clearly shows that increasing the average temperature suppresses the efficiency considerably, though the output power tends to increase. 
We understand that two factors contribute to this vulnerability in thermal efficiency. First, phonon transport increases with rising average temperature.  While some of the electron efficiency $\eta_{\text{el}}$ remains high (see Fig.~\ref{fig: eta vs power, (12, 4), double-bend}), the total efficiency $\eta$ considerably reduces.  The other factor is the finite-temperature effect smearing out the singularity of a Fano-type resonance. It prevents destructive quantum interference from enhancing thermal efficiency.

\subsection{Linear-response estimate of nonlinear thermoelectricity}
\label{sec:linear-response-estimate}

In retrospect, we have started by examining linear-response quantities to search for a nanostructure suitable for high nonlinear thermoelectricity. The significance of the linear-response estimate has already proved itself by comparing the linear and nonlinear power-efficiency diagrams (Fig.~\ref{fig: eta vs power, linear, double-bend} and Fig.~\ref{fig: eta vs power, T_bar is const, double-bend}).  To make such a direct comparison, we have found it crucial to normalize the thermal efficiency and output power and to use the average temperature (see also the argument in \cite{Azema14,Taniguchi20} for an appropriate choice of the temperature). In this section, we will examine it more closely and argue that we can assess the nonlinear performance of a heat engine reasonably well based on the linear response theory. 

%
\begin{figure}
  \centering
  \subfloat[Linear-response estimate\label{fig: double-bend eta_el, power, T_bar=4, eta=0}]{
    \includegraphics[width=0.49\linewidth, trim=20 30 10 0]{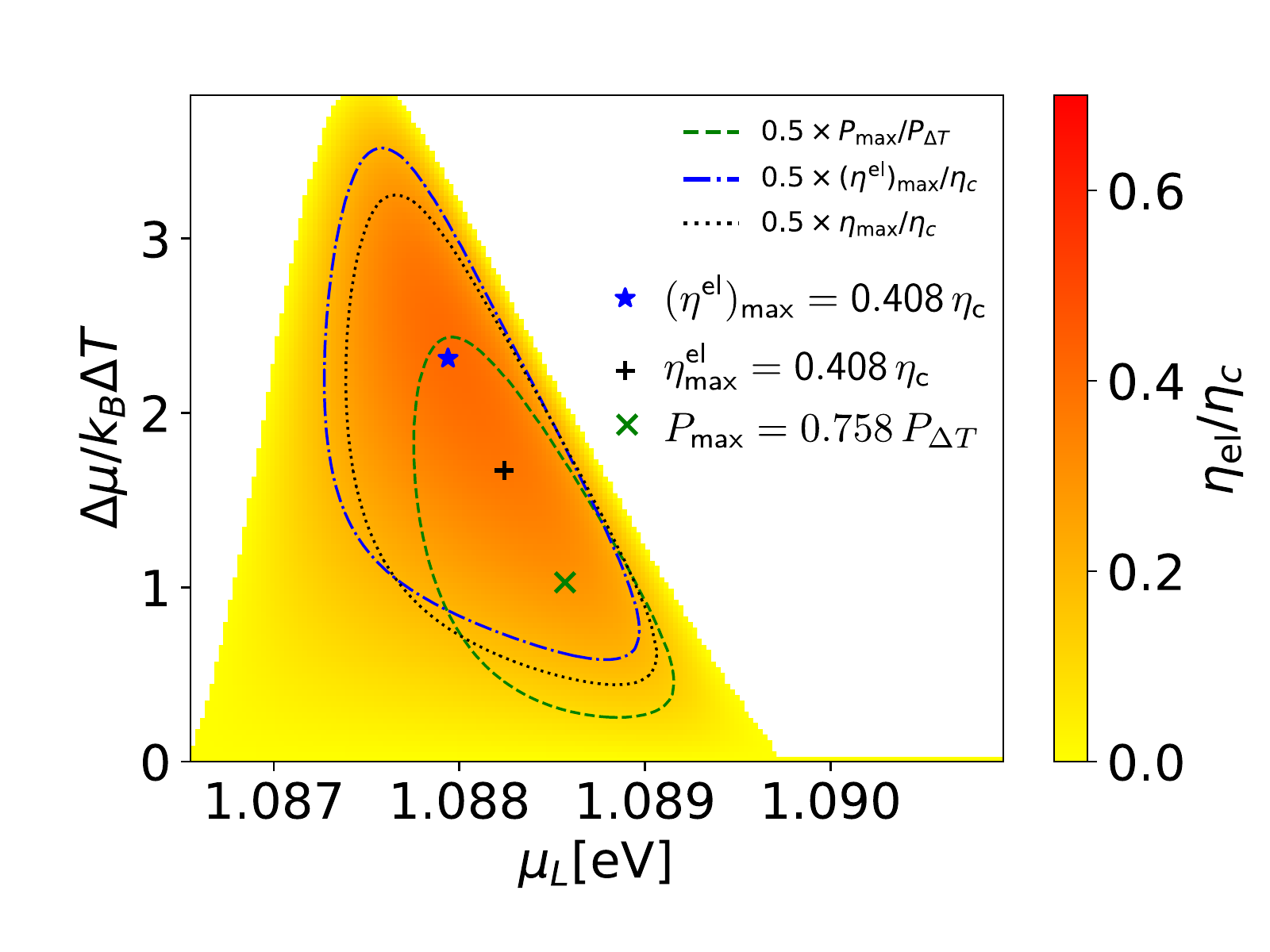}
  }
  \subfloat[Weak nonlinear regime\label{fig: double-bend eta_el, power, T_bar=4, eta=1/3}]{
    \includegraphics[width=0.49\linewidth, trim=20 30 10 0]{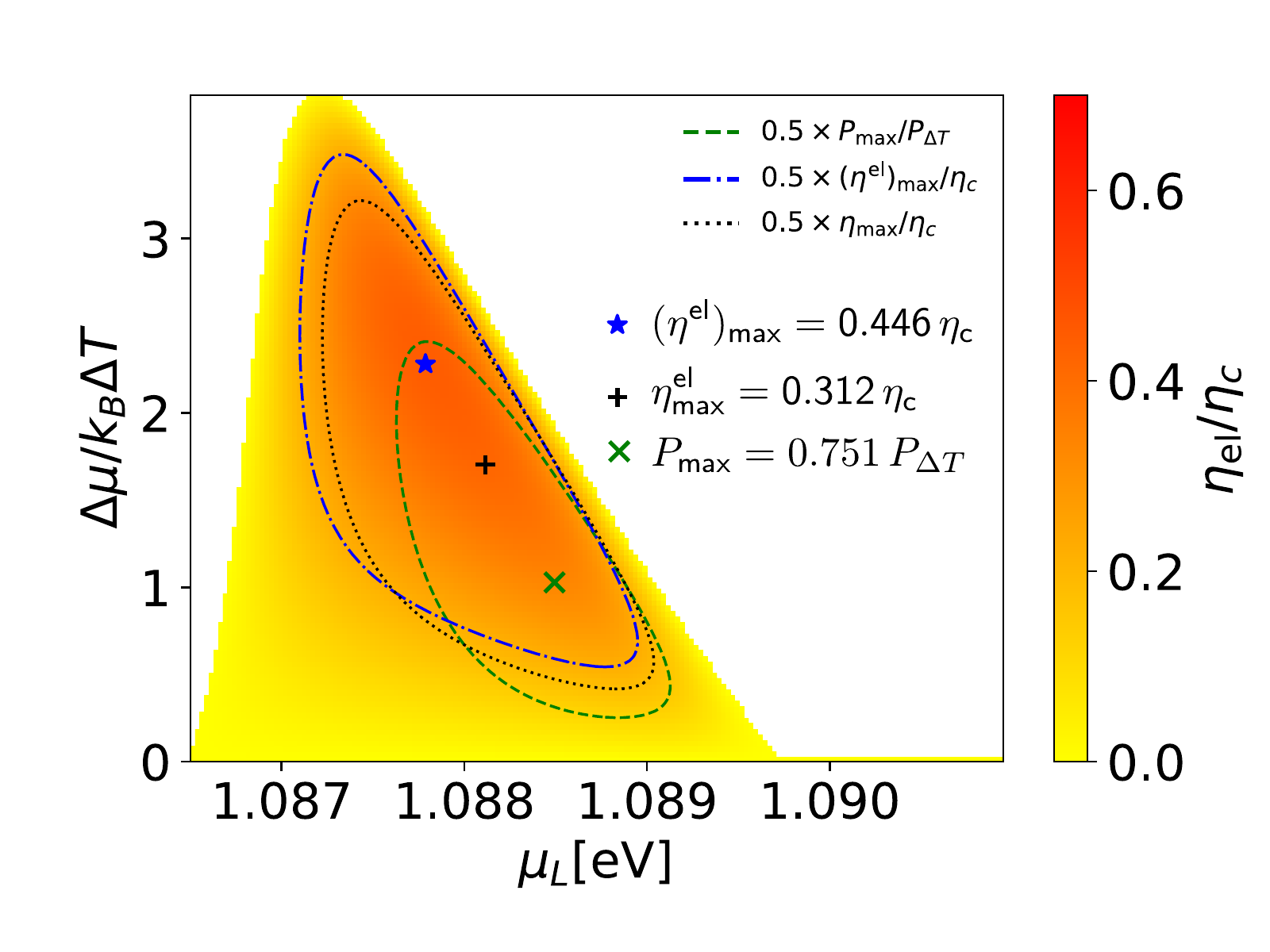}
  } \\
  \subfloat[Intermediate nonlinear regime\label{fig: double-bend eta_el, power, T_bar=4, eta=2/3}]{
    \includegraphics[width=0.49\linewidth, trim=20 30 10 30]{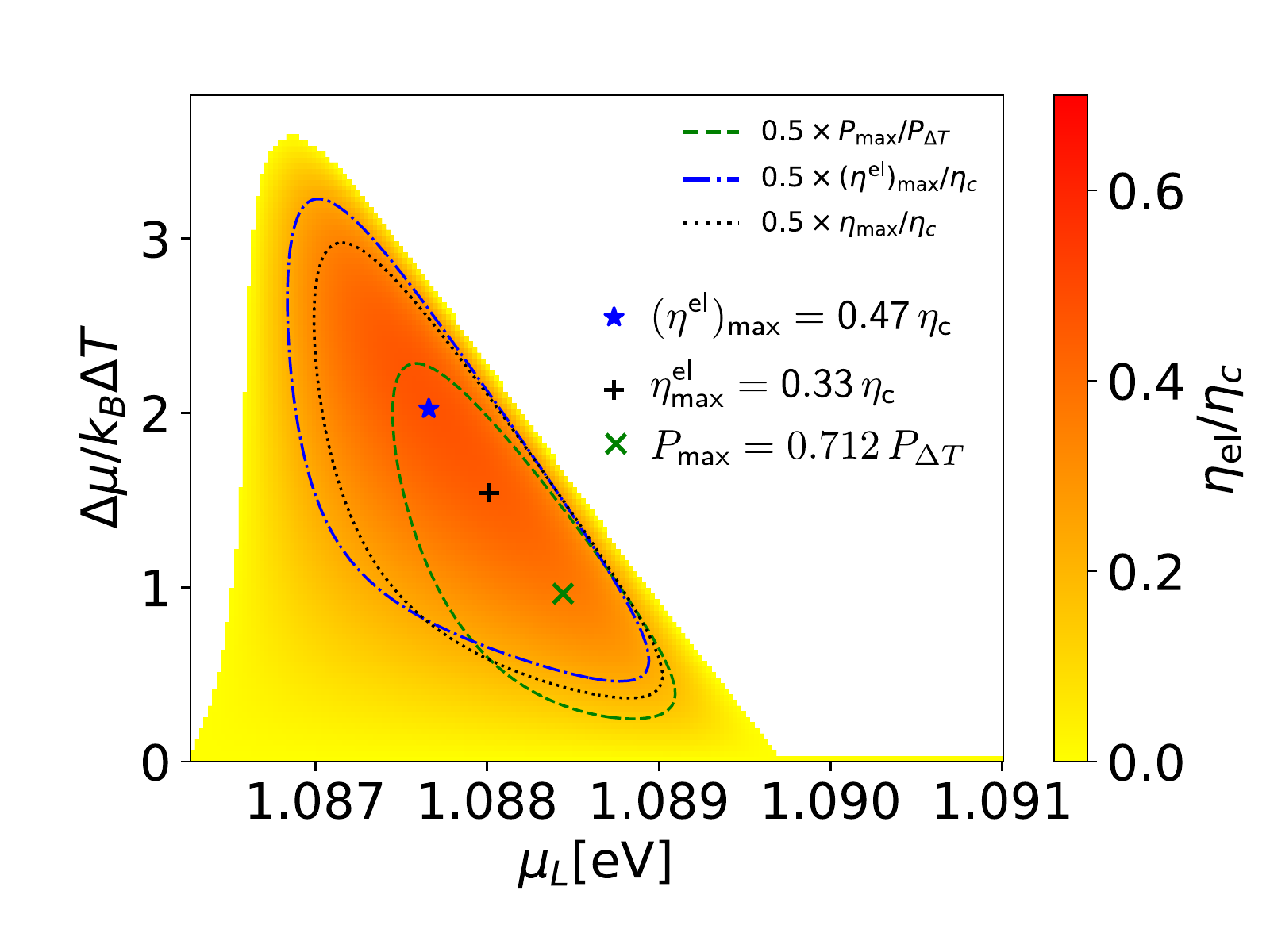}
  }
  \subfloat[Strong nonlinear regime\label{fig: double-bend eta_el, power, T_bar=4, eta=0.9}]{
    \includegraphics[width=0.49\linewidth, trim=20 30 10 30]{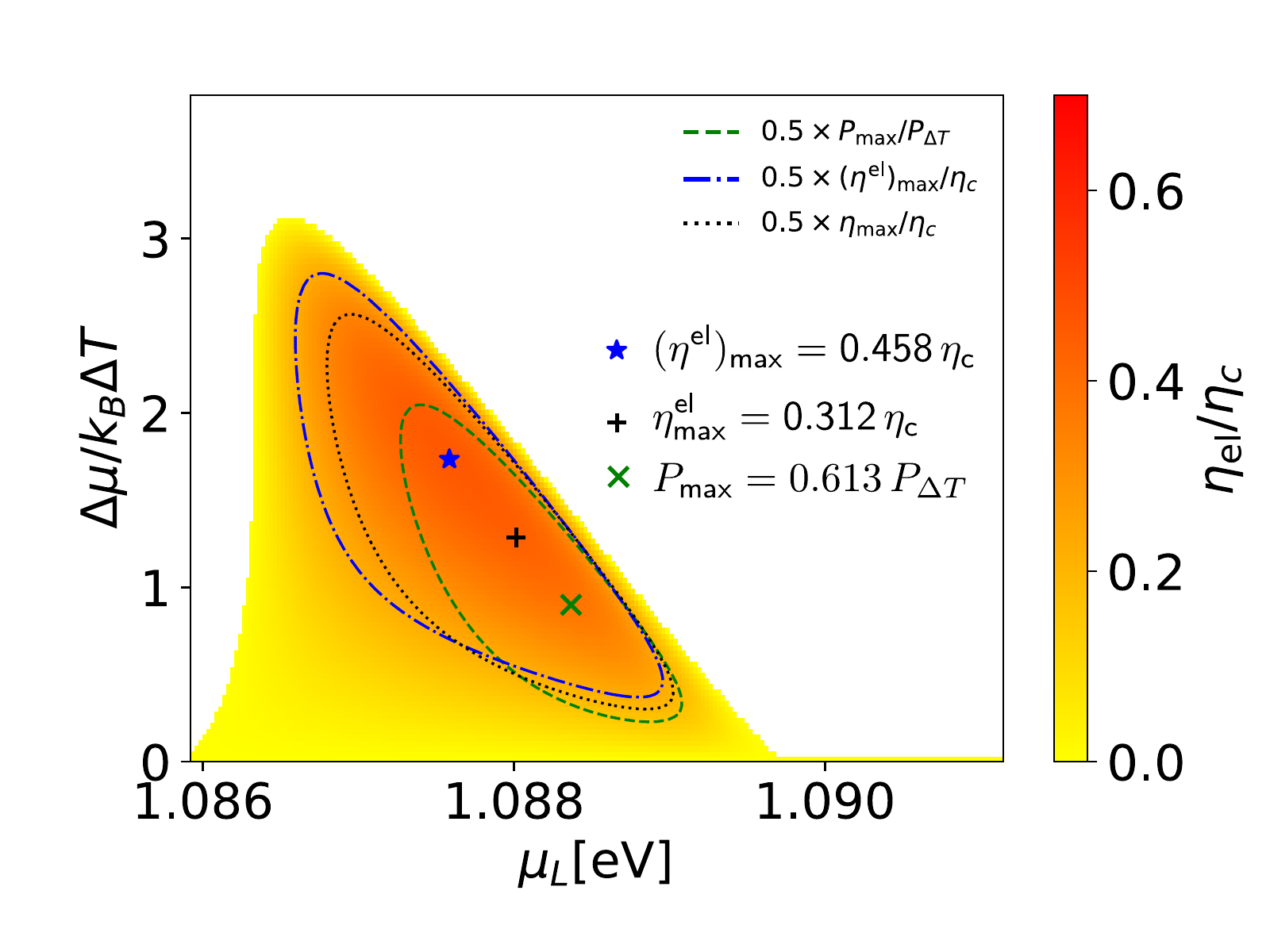}
  }
  \caption{The dependence of the normalized efficiency on the normalized bias voltage $\Delta \mu/k_{B}\Delta T$ and the gate voltage $\mu_{L}$. Linear-response result at (a) $T=\SI{4}{K}$ is compared with nonlinear-response regimes of (b)$\Delta T<\bar{T}$, (c) $\Delta T=\bar{T}$  and (d) $\Delta T>\bar{T}$. (b) $(T_{L},T_{R})=(\SI{4.8}{K},\SI{3.2}{K})$ at $\eta_{c}=1/3$, (c) $(T_{L},T_{R})=(\SI{6}{K},\SI{2}{K})$ at $\eta_{c}=2/3$ and (d) $(T_{L},T_{R})=(\SI{7.27}{K}, \SI{0.73}{K})$ at $\eta_{c}=0.9$. In each subfigure, half values of $\eta_{\rm max}$, $P_{\rm max}$ and $\eta_{\rm max}^{\rm el}$ are depicted respectively by black dotted, green dashed and blue dot-dashed lines.
  \label{fig:efficiency-by-mu-and-bias-4K}}
\end{figure}

In Fig.~\ref{fig:efficiency-by-mu-and-bias-4K}, we closely examine how linear and nonlinear efficiencies depend on the bias voltage and the chemical potential in the double-bent rhombus dot at $\epsilon_{g}=\SI{5.40}{eV}$ (with the Fano resonance peak at $\mu\approx\SI{1.08}{eV}$). 
%
%
We recall that the linear-response theory can provide an estimate of the normalized power $P/P_{\Delta T}$ and the electronic efficiency $\eta_{\text{el}}/\eta$ in the full range of bias voltage ratio $0\le\Delta \mu/\Delta \mu_{\text{stop}}\le 1$ (see the last paragraph of Sec.~\ref{sec:linear-response-theory}). 
Accordingly, normalizing the bias voltage $\Delta \mu$ by $k_{B}\Delta T$ enables us to make direct and detailed comparisons between linear and nonlinear-response results. 
We have prepared ``linear-response estimate'' at \SI{4}{K} in Fig.~\ref{fig: double-bend eta_el, power, T_bar=4, eta=0}, calculated entirely by linear-response quantities.  This result is compared with nonlinear responses of weakly nonlinear $\bar{T}>\Delta T$, intermediately nonlinear $\bar{T}=\Delta T$ and strongly nonlinear $\bar{T}<\Delta T$ results: $(T_{L},T_{R})=(\SI{4.8}{K},\SI{3.2}{K})$ at $\eta_{c}=1/3$ (Fig.~\ref{fig: double-bend eta_el, power, T_bar=4, eta=1/3}), 
$(T_{L},T_{R})=(\SI{6}{K},\SI{2}{K})$ at $\eta_{c}=2/3$ (Fig.~\ref{fig: double-bend eta_el, power, T_bar=4, eta=2/3}) and $(T_{L},T_{R})=(\SI{7.27}{K}, \SI{0.73}{K})$ at $\eta_{c}=0.9$ (Fig.~\ref{fig: double-bend eta_el, power, T_bar=4, eta=0.9}).  
First, Fig.~\ref{fig:efficiency-by-mu-and-bias-4K} confirms a considerable overlap between the region with high efficiency (inside of the black dotted line, $\eta>0.5\eta_{\max}$) and the one with high output power (inside the green dashed line, $P>0.5P_{\max}$).  
We also see the linear-response result (Fig.\ref{fig: double-bend eta_el, power, T_bar=4, eta=0}) capture the essence of nonlinear responses (Fig.~\ref{fig: double-bend eta_el, power, T_bar=4, eta=1/3}, \ref{fig: double-bend eta_el, power, T_bar=4, eta=2/3}, \ref{fig: double-bend eta_el, power, T_bar=4, eta=0.9}) quite well even in fully nonlinear regimes, while the stopping voltage gets increasingly suppressed by increasing the nonlinearity, especially for $\eta_{c}\approx 0.9$ (Fig.~\ref{fig: double-bend eta_el, power, T_bar=4, eta=0.9}).   
In addition, the linear-response estimate identifies the locations of optical parameters for achieving the highest efficiency and output power. Therefore, we can rely on the linear-response results to predict the nonlinear thermoelectric performance of a nanoscale heat engine. 
We remark that the present situation is quite different from graphene-superconductor and superconductor-superconductor tunnel junctions \cite{Bernazzani23,Germanese22,Germanese23}, where strong thermoelectricity appears only in the nonlinear regime with almost vanishing linear thermoelectricity.

\section{conclusion}
\label{sec:conclusion}

We have theoretically explored how to enhance linear and nonlinear thermoelectric performance in a nanoscale heat engine by making structural modifications on a graphene rhombus dot. After evaluating the phonon and electron transport in a linear-response model, we have identified a nanostructure suitable for high thermoelectricity. Modifying the junction bending angle suppresses phonon transport, and Fano-like asymmetric resonances provide high efficiency. We have found that adjusting a tunable local gate voltage on a double-bent graphene rhombus dot is an effective way to achieve high efficiency and output power, particularly at low temperatures ($T=\SI{4}{K}$). We have also demonstrated how normalized linear-response plotting helps us predict nonlinear thermoelectric performance reliably. We believe controlling quantum coherence is a powerful method when searching for better thermoelectric materials at the nanoscale.

\begin{acknowledgements}
This work has been partly supported by JSPS (JP) KAKENHI Grant No.~JP19K03682. The authors acknowledge fruitful discussions with S.~Okada and Y.~Tokura. 
\end{acknowledgements}



\bibliography{../ref}

\end{document}